%% file: main.tex
\documentclass[11pt]{article}

\usepackage[final]{acl}

\usepackage{times}
\usepackage{latexsym}
\usepackage[T1]{fontenc}
\usepackage[utf8]{inputenc}
\usepackage{microtype}
\usepackage{inconsolata}
\usepackage{graphicx}
\usepackage{booktabs}
\usepackage{tabularx}
\usepackage{amssymb}
\usepackage{array}
\usepackage{float}
\usepackage{titletoc}
\usepackage{algorithm}
\usepackage{algorithmic}
\usepackage{amsmath}
\usepackage{multirow}
\usepackage{tcolorbox}
\tcbuselibrary{listings, skins, breakable} 
\usepackage{enumitem}   
\usepackage{multicol}   

\title{From Script to Stage: Automating Experimental Design for Social Simulations with LLMs}

\author{
\textbf{Yuwei Guo\textsuperscript{1,2,3}},
\textbf{Zihan Zhao\textsuperscript{1,2,3}},
\textbf{Xiaowei Liu\textsuperscript{1,2,3}},
\textbf{Xiangning Yu\textsuperscript{1,2,3}},
\textbf{Qun Ma\textsuperscript{1,2,3}},
\textbf{Deyu Zhou\textsuperscript{4}},
\textbf{Xiao Xue\textsuperscript{1,2,3}\thanks{Corresponding author}}
\\[0.5em]
\textsuperscript{1}College of Intelligence and Computing, Tianjin University, Tianjin, China\\
\textsuperscript{2}Tianjin Key Laboratory of Healthy Habitat and Smart Technology, Tianjin, China\\
\textsuperscript{3}Laboratory of Computation and Analytics of Complex Management Systems, Tianjin University, Tianjin, China\\
\textsuperscript{4}School of Software, Shandong University, Shandong, China
\\[0.5em]
\texttt{\{2024244171, zhaozihan, xiaoweiliu, yxn9191, 1023244018, jzxuexiao\}@tju.edu.cn}, \\
\texttt{202220787@mail.sdu.edu.cn}
}

\begin{document}
\maketitle

\begin{abstract}
    \input{sections/00_abstract}
\end{abstract}

\input{sections/01_intro}
\input{sections/02_related_work}
\input{sections/03_method}
\input{sections/04_experiment}
\input{sections/05_conclusion}

\input{sections/06_limitations}
\input{sections/07_ethical_cons}
\input{sections/acknowledgments}

\bibliography{custom}

\clearpage
\appendix
\input{sections/08_appendix}

\end{document}

%% file: sections/00_abstract.tex
Multi-agent simulation based on LLMs has increasingly emerged as a new paradigm for exploring complex social phenomena and validating theoretical hypotheses. However, traditional experimental design in the social sciences relies heavily on interdisciplinary expert knowledge, involving cumbersome procedures and high technical barriers. While LLM-driven agents demonstrate broad prospects for designing experiments, their limitations regarding reliability and scientific rigor continue to significantly hinder their in-depth application in social science research. To address these challenges, this paper proposes \textbf{FSTS}, an automated framework for multi-agent experiment design based on script generation. Drawing on the concept of the ``Decision Theater,'' the framework deconstructs experimental design into three core phases: \textbf{S}cript \textbf{C}omposition, \textbf{S}cript \textbf{F}inalization, and \textbf{A}ctor \textbf{G}eneration. Tests across multiple scenarios indicate that the agents generated by this framework can enact the script within the ``experimental theater,'' reproducing results consistent with real-world situations. The proposal of FSTS not only effectively lowers the barrier for social science experimental design but also provides scientifically grounded decision support for policy-making.
\footnote{The code repository for this project is publicly available at \url{https://github.com/RisingDate/FSTS}.}

%% file: sections/01_intro.tex
\section{Introduction}

In recent years, the rapid development of artificial intelligence technology, especially the widespread application of Large Language Models (LLMs), has driven a surge of research interest in ``AI for Social Science\cite{xu2024ai}.'' AI tools can not only alleviate the workload of researchers but also provide intelligent support in stages such as experimental design~\cite{xue2024computational}, variable selection, and scheme optimization.

In social science, \textbf{computational experiments}~\citep{xue2021computational, xue2024computational2} are often viewed as a ``third paradigm'' for studying complex systems in controllable virtual environments, yet adoption is limited by technical and interdisciplinary barriers. Decision-Theater-style approaches improve rigor through structured, participatory workflows~\citep{wolf2023decision} but require expert facilitation and substantial coordination, which hinders scalability~\cite{jaeger2026decision}. Meanwhile, LLM-based agents offer flexibility~\cite{tran2025multi} but are unreliable for rigorous experimental design due to hallucinations and limited verifiability~\cite{qin2023chatgpt,  messeri2024artificial, huang2025survey, chelli2024hallucination}.

Recent LLM-driven social simulation systems and environments (e.g., Smallville~\cite{park2023generative, xue2023chatgpt}, SOTOPIA~\cite{zhou2023sotopia,zhou2025sotopia}, and AgentSociety~\citep{piao2025agentsociety}) demonstrate promising progress in scalable interaction and evaluation. 
However, they largely leave open how to systematically compile high-level research requirements into standardized, inspectable, and executable experimental scripts with explicit control over variables, interventions, and evaluation criteria \cite{gao2024large, xue2021soa}.

A central bottleneck is the generation of \textbf{experimental scripts}: structured, executable specifications that translate high-level research questions into concrete experimental settings (e.g., environments, variables, interventions, and evaluation criteria). Despite recent progress, generating scripts that satisfy user requirements remains challenging for three reasons:
\begin{itemize}
    \item \textbf{Standardized yet expressive script specification.} Scripts must be machine-executable and well-structured, while remaining expressive enough to capture diverse evolutionary trajectories and plausible social dynamics~\cite{liu2024llms, liu2024we}.
    \item \textbf{Combinatorial complexity in experimental design.} Realistic social systems exhibit multi-factor dependencies, creating a trade-off between combinatorial explosion (an intractably large design space) and oversimplification (loss of validity).
    \item \textbf{Limited fidelity of agent modeling.} Many existing frameworks emphasize dialogue- or task-oriented behaviors, while richer agent attributes, relational structures, and mechanism-level assumptions are often under-modeled.
\end{itemize}

To address these challenges, we propose \textbf{FSTS}, an automated framework for experimental design for artificial-society simulation and reasoning. Inspired by film production, FSTS casts experimental design as a three-stage pipeline: (1) \textbf{Script Composition}, where a Screenwriter Agent generates multiple candidate experimental schemes from user requirements; (2) \textbf{Script Finalization}, where a Director Agent evaluates candidates along multiple dimensions and selects a final scheme; and (3) \textbf{Actor Generation}, where an Actor Factory instantiates experimental agents with attributes and relationship networks specified by the script. By introducing domain-specialized LLM agents and explicit intermediate artifacts at each stage, FSTS systematizes and automates the experimental design workflow while maintaining interpretability and user control.

The contributions of this paper are threefold:
\begin{itemize}
    \item We introduce \textbf{FSTS}, an LLM-centric framework that lowers the barrier to designing computational experiments for artificial-society simulation by bridging high-level research intent and executable experimental scripts.
    \item We propose a \textbf{film-inspired multi-agent workflow} (Screenwriter--Director--Actor Factory) that separates generation, critique/selection, and instantiation, improving the structure and reliability of script generation.
    \item We conduct an \textbf{empirical evaluation} with quantitative metrics, human judgments, and ablation studies, showing that FSTS produces higher-quality experimental scripts and agents than strong baselines.
\end{itemize}

%% file: sections/02_related_work.tex
\begin{figure*}[t]
\centering
\includegraphics[width=1\textwidth]{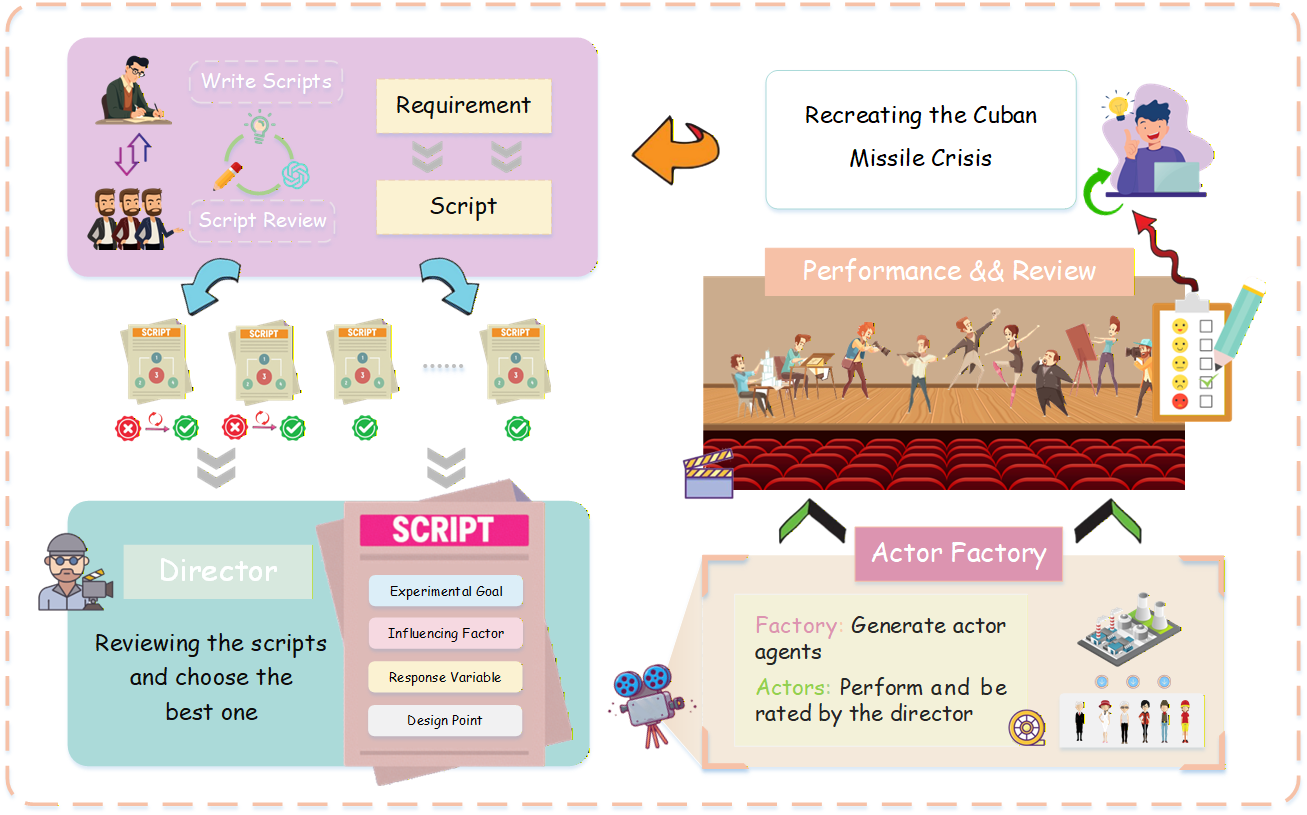}
\caption{\textbf{Overview of FSTS.}
First, the Screenwriter Agent generates multiple candidate scripts. Next, the Director Agent reviews each script and selects the final version. Subsequently, the Actor Factory generates Actor Agents to perform on the stage, and the performance results are fed back to the user for experimental optimization.}
\label{fig:framework}
\end{figure*}

\section{Related Work}
This section reviews related work on computational experiments. We focus on two lines of research that enable automated experimental design: requirement analysis methods~\cite{casalicchio2024ai, santos2024we} and emerging LLM-based multi-agent frameworks~\cite{chen2023autoagents, hong2024metagpt, yang2023auto}.

\subsection{Traditional Techniques}
Before the rise of computational social science and Large Language Models, decision-making and modeling in social science largely relied on traditional simulation- and facilitation-based approaches~\cite{xue2023computational2}.

\textbf{Computational experiments}~\cite{wang2004artificial, wang2004computational, xiao2023putational4, xue2023computational5} use computer modeling and simulation to quantitatively analyze complex social systems in a virtual environment, providing a technical basis for modeling and deduction~\cite{peng2023computational}. Built on visualization and simulation technologies, the \textbf{Decision Theater}~\cite{wolf2023decision} offers an integrated software--hardware setting that interactively presents decision schemes to support collective decision-making.

Compared with traditional Agent-Based Modeling (ABM)~\cite{macal2009agent, macal2005tutorial}, the Decision Theater combines large-screen visualization with roundtable deliberation, improving cognitive consistency and decision effectiveness. Nevertheless, it remains limited by (i) high professional requirements and cross-domain expert participation, (ii) dependence on on-site facilitation with complex and costly procedures, and (iii) persistent ``Big Data--Small Data'' issues, a relatively single decision-making paradigm, and limited human--machine collaboration.

\subsection{The Era of LLM agent}
LLM-based agent simulation has become a major direction~\cite{piao2025agentsociety, gurcan2024llm,tang2025gensim, ma2024computational}. \autoref{tab:framework_comparison} compares currently popular agent frameworks. Relative to ABM, it lowers modeling barriers via natural language, enables richer planning and coordination, and introduces stochasticity. However, it remains bottlenecked by prompt-heavy generation without active validation and by hallucinations and bias, which undermine rigor and credibility~\cite{mao2025agent,zhang2025socioverseworldmodelsocial}.

\subsection{Limitations of Existing Approaches}
Traditional computational experiments and Decision-Theater-style workflows are rigorous but costly and hard to scale, whereas LLM-driven approaches are accessible but difficult to trust for high-stakes design. To reconcile this trade-off, we propose \textbf{FSTS}, a \textbf{Screenwriter--Director--Actor} framework that closes the loop from user requirements to experimental enactment while balancing rigor, credibility, and usability.

%% file: sections/03_method.tex
\section{Methodology}
Drawing inspiration from the film production workflow, we divide the process of automated experimental design into three key stages: \textbf{Script Composition}, \textbf{Script Finalization}, and \textbf{Actor Generation}. As shown in \autoref{fig:framework}, the framework starts with user requirements and finally generates Actor Agents capable of performing in the designated theater.

\subsection{Script Composition}
\begin{figure}[t]
    \centering
    \includegraphics[width=1\linewidth]{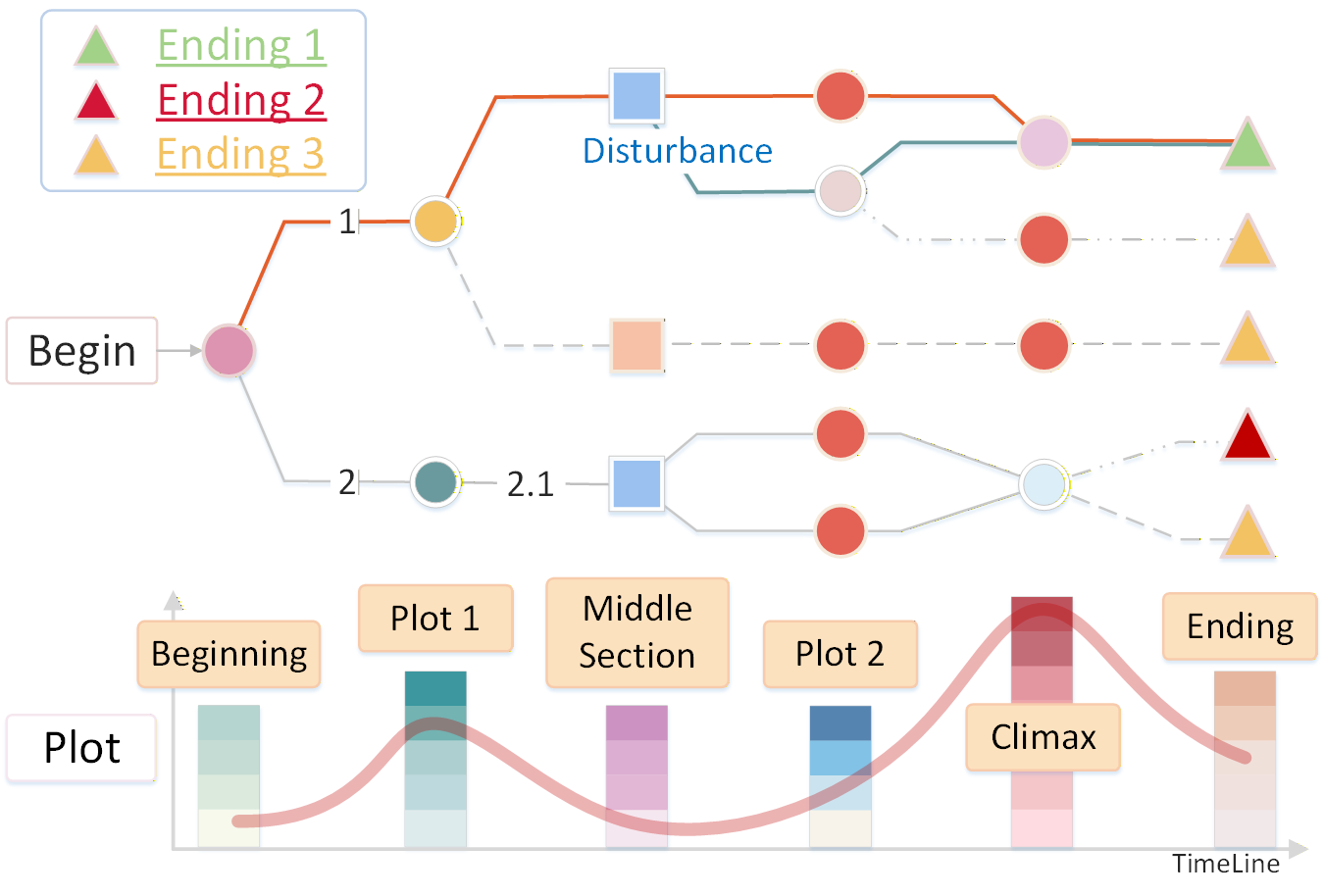}
    \caption{\textbf{Dynamic Mechanism of Script Evolution.} Non-linear bifurcation triggered by key decisions and the ultimate convergence of states.}
    \label{fig:storyline}
\end{figure}

The design of the experimental script aims to capture the non-linear evolutionary characteristics of real-world events. As shown in \autoref{fig:storyline}, the progression of events is replete with bifurcations and turning points: the choices made by agents at critical nodes determine subsequent trajectories, triggering distinct evolutionary branches and leading to diverse outcomes. Furthermore, the experimental script must be capable of simulating complex causal chains, allowing different evolutionary paths to reconverge at similar or identical final states. The precise modeling of this path dependence and system convergence constitutes a core requirement of this method.

Drawing inspiration from the industrial pipeline mechanism~\cite{sivasankaran2014literature} characterized by task decomposition and hierarchical collaboration, this paper proposes a hierarchical experiment script generation architecture. This architecture deconstructs complex experimental design tasks into standardized sub-modules, achieving the mapping from macro goals to micro parameters through multi-agent collaboration. We formally define the Experiment Script as a 5-tuple:

\begin{equation}
  S=<G, I, R, D, L>  
\end{equation}

Where $G$ represents the Experiment Goal, $I$ represents Influencing Factors, $R$ represents Response Variables, $D$ denotes parameterized Design Points, and $L$ represents the Storyline that runs through the experimental logic.

To improve the rationality and accuracy of the LLM output, the framework introduces a dual constraint mechanism on both the input and output sides:

\begin{itemize}
    \item \textbf{Input Side: Structured Guidance and Step-wise Generation.} We constrain the model's inference space using Domain Knowledge Templates, requiring user input to include three elements: research goal, core variables, and target objects. Simultaneously, a Step-wise Generation strategy is adopted to decompose the construction process of script $S$ into progressive sub-tasks targeting $G, I, R, D$ and $L$, ensuring local semantic accuracy.
    \item \textbf{Output Side: Multi-View Integration and Format Constraints.} A Multi-View Generation mechanism is introduced, where Screenwriter Agents generate candidate schemes from different perspectives (e.g., goal-oriented vs. process-oriented), and the optimal solution is selected via weighted voting. Additionally, strict JSON Schema Constraints are applied to ensure that the generated scripts can be directly parsed and executed by downstream simulation systems.
\end{itemize}

\begin{figure}[H]
    \centering
    \includegraphics[width=1\linewidth]{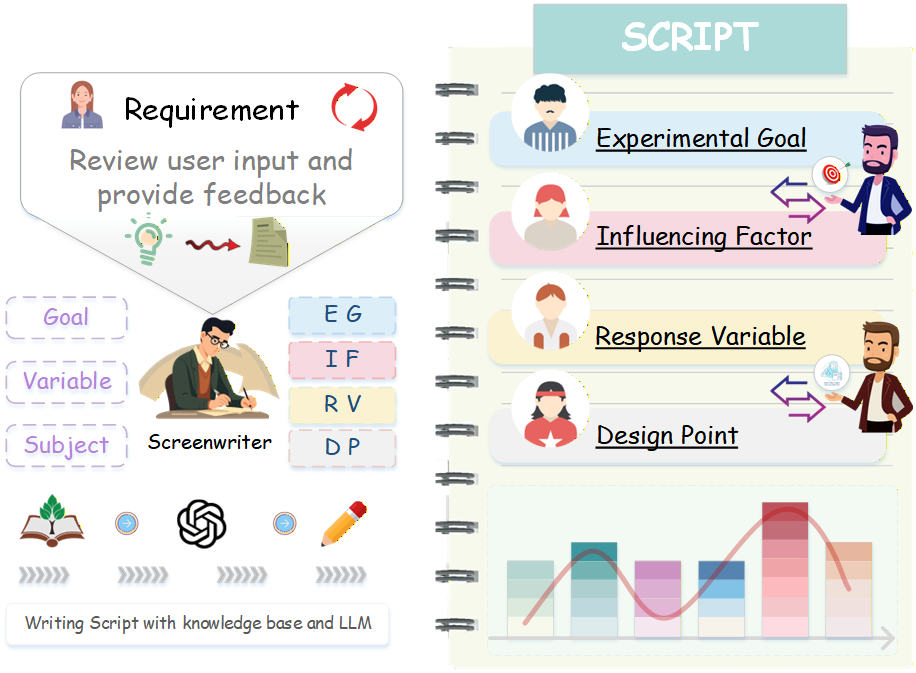}
    \caption{\textbf{Schematic Diagram of Script Generation.} The Screenwriter composes scripts by leveraging a knowledge base.}
    \label{fig:generation_script}
\end{figure}

As illustrated in \autoref{fig:generation_script}, this framework establishes a collaborative \textbf{Screenwriter-Director} system for experimental script generation. At the generation end, the Screenwriter Agent (GPT-4o)~\cite{openai2024gpt4o} is responsible for parsing user requirements and integrating domain knowledge to construct a five-dimensional preliminary script, encompassing elements such as experimental goals and storylines. For the critical Design Points ($D$), we adopt Design of Experiments (DOE) methods to optimize complexity~\cite{xue2024computational}, aiming to obtain complete experimental conclusions with as few experimental schemes as possible. At the monitoring end, four Director Agents (GPT-5 mini)~\cite{openai2025gpt5mini} are deployed to conduct itemized reviews focusing on the rationality and standardization of the script. The system employs a Cascading Validation mechanism: the review process adheres to a strict sequence, where the workflow transitions to the subsequent stage only after the preceding module has passed validation. Once a defect is detected, a feedback closed-loop is immediately triggered, guiding the Screenwriter Agent to perform targeted revisions on specific modules based on the Director's feedback.

\subsection{Script Finalization}
In the Script Composition phase, the Screenwriter Agent generates multiple candidate experimental scripts from different perspectives based on user requirements. Obtaining candidate scripts requires the further selection of a unique experimental plan for execution. Manual screening is not only costly and prone to subjective bias but also contradicts the objective of automated experimental design proposed in this paper. Therefore, we adopt the ``LLM as Judge'' approach to automatically evaluate and select candidate scripts.

Existing research indicates that LLMs may exhibit disadvantages in judging tasks, such as position bias, verbosity bias, self-enhancement bias, and limited reasoning capabilities~\cite{zheng2023judging}, which affect the stability of evaluation results. To address this, we referenced the work of Zheng et al. and introduced GPT-5 mini, Qwen-235b, and DeepSeek-V3.2, which are models more capable than the Screenwriter Agent, to form a multi-model voting committee for scoring the results during the evaluation phase. We designed a two-stage script finalization mechanism combining Single-Answer Scoring and Pairwise Comparison to enhance the reliability and rationality of the selection process.

\textbf{Single-Answer Scoring.} In the first stage, the LLM Judge independently scores all candidate scripts generated by the Screenwriter Agent on a 100-point scale. The top 4 scripts with the highest scores are retained for the next stage.
    
During scoring, we require the judge to follow Nigel Gilbert’s core ABM criterion~\cite{gilbert2019agent}: experiments should demonstrate a mechanism’s \emph{explanatory power}, not merely that a model can run. Accordingly, the LLM Judge evaluates scripts along Gilbert-aligned dimensions—\emph{theory-driven design}, \emph{progressive complexity}, \emph{controlled and randomized repetition}, \emph{interpretability}, and \emph{robustness}—to ensure methodological compliance with computational social science experiments.

\textbf{Pairwise Comparison.} In the second stage, we perform pairwise comparisons on the 4 candidate scripts selected from the preliminary screening. In each comparison, the LLM Judge must explicitly state ``which script is better'' and provide a brief rationale. We ultimately determine the optimal plan through a points-based system.

To mitigate potential position bias in LLM comparison tasks, we swap the presentation order of the two scripts in each pair and conduct two independent evaluations. If the results of the two comparisons are inconsistent, we introduce a new LLM Judge to directly finalize the better one.

Furthermore, if non-transitive preferences occur during pairwise comparisons (e.g., A is better than B, B is better than C, but C is better than A), an additional evaluation round is triggered. The LLM Judge will strictly rank the relevant scripts to resolve potential conflicts and improve the consistency of the final decision.

By combining single-answer scoring with pairwise comparison and explicitly controlling for known LLM weaknesses in model selection and process design, this method maintains the degree of automation while improving the stability and interpretability of the script finalization phase. It provides reliable input for subsequent simulation experiments. Additionally, we conducted two supplementary experiments. First, we compared our method with approaches using only single-answer scoring or only pairwise comparison to screen experimental scripts, and calculated the average runtime, result entropy~\cite{yu2024beyond,yu2025unlocking}, and aggregation degree (Top-1 Confidence) of these methods using different LLM kernels. The specific results are shown in \autoref{tab:finalization_method_comparison}. Second, we supplemented a rigorous human alignment study, in which we invited sociologists and computer scientists to finalize the scripts and compared their selections with the results chosen by the LLM acting as the Director. The results are presented in \autoref{tab:krippendorff_alpha}, and the detailed experimental setup is provided in the Appendix \ref{subsec:human_alignment_study}.

\begin{table}[htbp]
    \centering
    \caption{Comparison of evaluation methods.}
    \label{tab:finalization_method_comparison}
    \small
    \begin{tabularx}{0.48\textwidth}{cccc}
    \toprule
        \textbf{Method} & \textbf{Runtime} & \textbf{Entropy} &\textbf{Top-1 Confidence}\\
        \midrule
        Single      & 253 & 1.295 & 0.60\\
        Pairwise    & 2657 & 1.000 & 0.50\\
        \textbf{Both}        & \textbf{613} & \textbf{0.722} & \textbf{0.80}\\
        \bottomrule
    \end{tabularx}
\end{table}

\subsection{Actor Generation}
To address the bottleneck where existing frameworks are limited to shallow interactions and lack deep causal deduction, the Actor Factory aims to build cognitive entities capable of \textbf{deductive reasoning}. Through a modular injection mechanism, the system transforms general-purpose models into specialized Agents adapted to complex scenarios. Its cognitive architecture is supported by four pillars:

\textbf{Domain Knowledge Injection:} To break the limitations of general common sense, the system dynamically loads professional knowledge bases. This enables the Agent Factory to perform deep deduction based on professional causal chains, achieving expert-level decision-making mechanisms.

\textbf{State and Emotion Mapping:} By introducing dynamic affective computing, environmental stimuli are mapped to psychological states in real-time. This mechanism endows Actor Agents with Bounded Rationality, enabling them to simulate authentic human cognitive biases under stress or panic, thereby significantly enhancing the ecological validity of decisions.

\textbf{Adaptive Goal Planning:} Distinct from single-step task execution, Actor Agents integrate reinforcement learning feedback. They can dynamically adjust terminal goals and strategic priorities in response to environmental evolution~\cite{yu2025causal}.

\textbf{Heterogeneous Cognitive Customization:} Based on an attribute orthogonalization strategy, the system establishes differentiated cognitive stances for Agents. Agents with different backgrounds generate divergent reasoning paths for the same event; this micro-level logic bifurcation is the prerequisite for the emergence of macro-level social complexity.

Furthermore, we introduce a \textbf{Casting Director} to execute blocking validation. Simulating an adversarial perspective, this module subjects generated Agents to logical stress testing. This ensures they possess substantive reasoning self-consistency.

\subsection{Script-Guided Performance}
Algorithm~\ref{alg:enactment} summarizes the script-guided enactment procedure. Upon the finalization of the script and the deployment of actors, the system formally enters the enactment phase. Within discrete time steps, customized Actor Agents conduct autonomous deduction based on the script's storyline. Relying on the cognitive architecture described in previous section, agents strictly adhere to the \textbf{Perception-Reasoning-Decision-Action} closed-loop mechanism for interaction, thereby driving the dynamic evolution of the simulation environment.

During this process, the Director Agent assumes the function of runtime supervision. It monitors the logical consistency between the simulation trajectory and the script's storyline in real-time in the background. It is also responsible for tracking individual behaviors and collecting data at key nodes, ultimately generating response variable records. This realizes the closed-loop transformation from script design to experimental data.

After the experiment concludes, the Director Agent triggers the feedback mechanism, transmitting the experimental results back to the user and the Screenwriter Agent. This feedback signal serves as the basis for driving the revision of user requirements and the iterative optimization of the script scheme, ensuring that the experimental design continuously approaches the research goals. 


\begin{algorithm}[h]
\caption{Script-Guided Enactment Process}
\label{alg:enactment}
\begin{algorithmic}[1]
\REQUIRE Experiment Script $S = \langle G, I, R, D, L \rangle$, Actor Set $\mathcal{A}$
\ENSURE Experiment Response Log $\mathcal{L}$

\STATE \textbf{Phase 1: Initialization}
\STATE Deploy Actors $\mathcal{A}$ into the social network

\STATE \textbf{Phase 2: Enactment Loop}
\FOR{$t = 1$ to $T_{max}$}
    \STATE \COMMENT{\textbf{Director Supervision}}
    \IF{Deviation from Storyline $S.L$ detected}
        \STATE Inject \textit{Correction Event} to align trajectory
    \ENDIF

    \STATE \COMMENT{\textbf{Actor Cognitive Cycle}}
    \FOR{each agent $a \in \mathcal{A}$}
        \STATE $obs \leftarrow \text{Perceive}(E, a.P, a.I)$
        \STATE $reasoning \leftarrow \text{Deduce}(obs, a.K)$ \COMMENT{Knowledge-based Reasoning}
        \STATE $decision \leftarrow \text{Plan}(reasoning, a.G)$ \COMMENT{Goal-driven Planning}
        \STATE $action \leftarrow \text{Act}(decision, a.I)$ \COMMENT{Modulated by Affect}
        \STATE Update Environment $E$ with $action$
    \ENDFOR

    \STATE \COMMENT{\textbf{Data Collection}}
    \STATE Record key node states into Log $\mathcal{L}$ based on $S.R$
\ENDFOR

\RETURN $\mathcal{L}$
\end{algorithmic}
\end{algorithm}

%% file: sections/04_experiment.tex
\section{Experiments}
To evaluate the capabilities and performance of the framework in automating the experimental design process, we conducted extensive experiments on benchmark tasks.

\subsection{Experimental Setup}

\begin{figure}[H]
    \centering
    \includegraphics[width=1\linewidth]{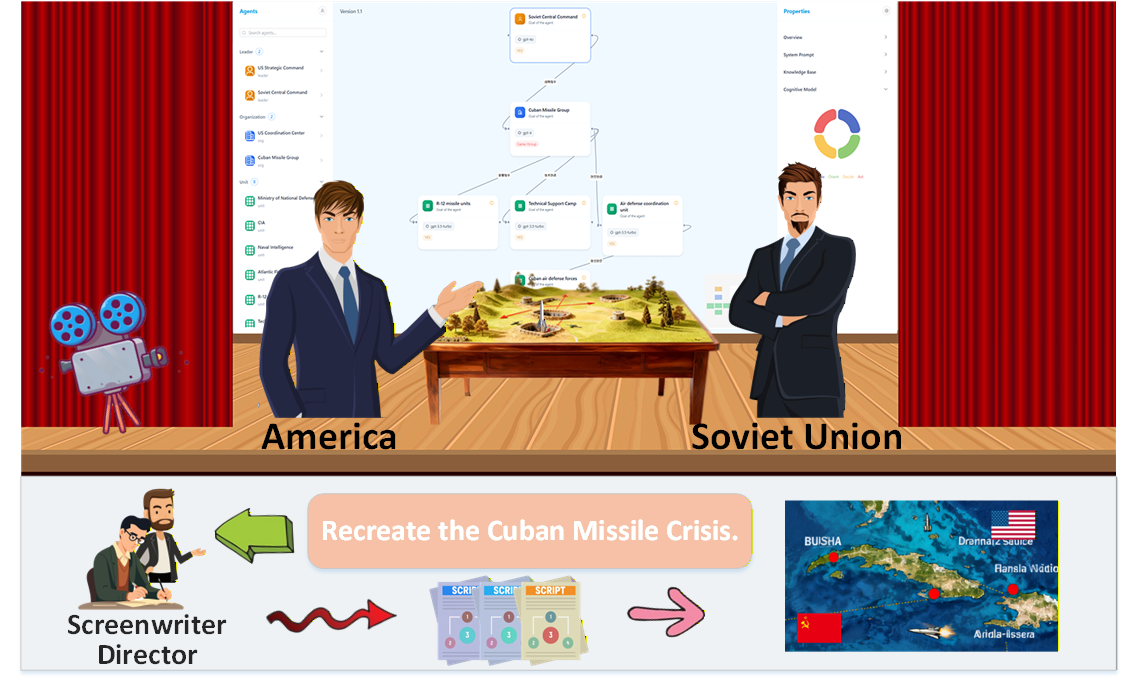}
    \caption{Schematic diagram of the experimental scenario.}
    \label{fig:placeholder}
\end{figure}

\textbf{Experimental Scenario}. To verify the effectiveness of the framework, we selected the 13-day Cuban Missile Crisis strategic game scenario. Involving two state agents, the USA and the USSR, this historical event offers distinct advantages: rich data availability, multi-level modeling capabilities, a clear timeline with critical decision nodes, and traceable results.

\textbf{Model Selection}. We employed GPT-4o~\citep{openai2024gpt4o} and GPT-5 mini~\citep{openai2025gpt5mini} as the core models for script generation and monitoring. Detailed parameter settings are provided in Appendix \ref{appendix-sec:hyperparameter}.

\begin{figure}[t]
    \centering
    \includegraphics[width=1\linewidth]{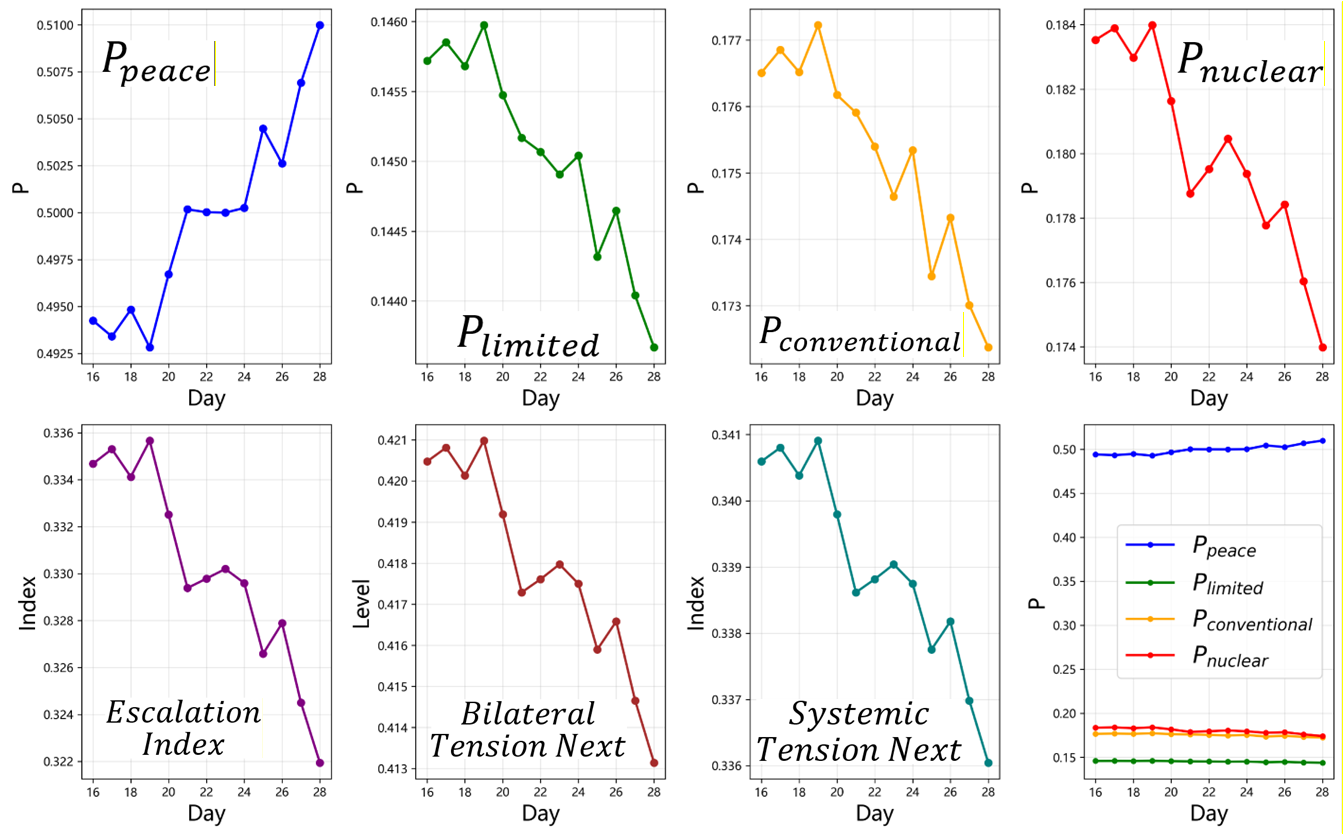}
    \caption{Result of scenario reproduction experiment.}
    \label{fig:data1}
\end{figure}

\textbf{System Input}. Guided by the feedback from experimental results, we continuously adjusted the system inputs. The initial user input was: ``I want to reproduce the Cuban Missile Crisis.'' The final improvements to the experimental requirements are shown in \autoref{fig:req_change}.

\textbf{Result Evaluation}. For result evaluation, we synthesized two primary criteria: First, the degree of alignment between actions taken by Actor Agents at critical time nodes and those taken by historical states (or corresponding leaders); Second, the consistency of the simulation's final outcome with historical reality. The historical events referenced in the experiment are detailed in the \autoref{tab:cuban_crisis_timeline_en}.

When comparing the behavioral decisions of agents in the experiment with the actions taken by historical national leaders, we employed Sentence-BERT~\cite{reimers-2019-sentence-bert} and GPT-5 mini respectively to assess the semantic similarity between historical outcomes and experimental simulation results.

\subsection{Main Results}
After iterative Screenwriter–Director interactions, the system produced 10 candidate scripts. Using six criteria, the Chief Director selected Script~2 as the final script (83.5), which specifies 29 influencing factors, 4 response variables, and 12 experimental design points.

The response variables in the script include: \textbf{war event outcome probabilities}, \textbf{escalation index}, \textbf{bilateral tension next}, and \textbf{systemic tension index}. Specifically, \textbf{war event outcome probabilities} is categorized into four outcomes: $[P_{peace}, P_{limited}, P_{conventional}, P_{nuclear}]$, representing the probabilities of peace, limited conflict, conventional war, and nuclear war, respectively. The changes in these variables during the 13-day simulation are illustrated in \autoref{fig:data1}. It can be observed that the probability of peace shows an overall upward trend, while the probability of war shows a downward trend. Notably, the probability of war rose briefly on the 19th and between the 23rd and 24th, coinciding with the scheduled time points of inject events.

Evaluated using the Sentence-BERT model, the semantic similarity between the decisions made by the script-generated Actor Agents and the actions taken by historical national leaders was 53.50, whereas it reached 72.50 using GPT-5 mini. The final outcome of the event simulation was ``peaceful resolution, but tense relations,'' which is consistent with historical reality.

\begin{table}[H]
\caption{Comparison of experimental design methods and full factorial methods.}
\label{tab:DoE-result}
\small
\centering
\renewcommand\arraystretch{1.2} 

\resizebox{0.5\textwidth}{!}{
    \begin{tabular}{cccccccc} 
    \toprule
    \multirow{2}{*}{Method} & 
    \multirow{2}{*}{Num($I$)} & 
    \multirow{2}{*}{Num($R$)} & 
    \multirow{2}{*}{Num(Exp)} &
    \multicolumn{4}{c}{Result} \\
    \cmidrule(lr){5-8}
    
    & & & & $p_1$ & $p_2$ & $p_3$ & $p_4$\\
    \midrule
    
    Total Factor    &5  &2  &243    &0.091  &0.708  &0.123  &0.078 \\
    \textbf{Ours}   &5  &2  &27     &0.148  &0.593  &0.148  &0.111 \\
    \bottomrule
    \end{tabular}
}
\end{table}

\subsection{Experimental simplification}

\begin{table*}[t]
    \centering
    \caption{Comparison of different order distribution methods.}
    \label{tab:platform_performance}
    \small
    \renewcommand\arraystretch{1}
    
    \resizebox{\textwidth}{!}{%
        \begin{tabular}{cccccc}
            \toprule
            \textbf{Distribution Mode} & \textbf{H-M Ratio} & \textbf{Runtime (s)} & \textbf{Orders} & \textbf{Total Platform Cost} & \textbf{Platform Efficiency} \\
            \midrule
            Normal Distribution & 1:2 & 165.33 & 124.33 & 56742.33 & 38731.67 \\
            Normal Distribution & 1:1 & 123.67 & 121.67 & 57866.33 & 36206.67 \\
            Normal Distribution & 2:1 & 116.00 & 132.00 & 66783.00 & 39061.00 \\
            \midrule
            Uniform Distribution & 1:2 & 144.33 & 110.67 & 50386.33 & 32892.67 \\
            Uniform Distribution & 1:1 & 116.67 & 131.00 & 59329.00 & 35345.00 \\
            Uniform Distribution & 2:1 & 109.00 & 112.33 & 56568.33 & 26679.67 \\
            \bottomrule
        \end{tabular}%
    }
\end{table*}

\begin{table*}[t]
    \centering
    \caption{Comparison of different human-machine difficulties and ratios.}
    \label{tab:difficulty_ratio_comparison}
    
    \renewcommand\arraystretch{1}
    
    \resizebox{\textwidth}{!}{%
        \begin{tabular}{cccccc}
            \toprule
            \textbf{Order Difficulty} & \textbf{H-M Ratio} & \textbf{Runtime (s)} & \textbf{Orders} & \textbf{Total Platform Cost} & \textbf{Platform Efficiency} \\
            \midrule
            2 & 1:2 & 165.33 & 124.33 & 56742.33 & 38731.67 \\
            1 & 1:2 & 120.83 & 121.00 & 38845.00 & 24704.00 \\
            3 & 1:2 & 215.33 & 116.00 & 78906.67 & 53325.33 \\
            3 & 3:1 & 136.67 & 126.67 & 96494.33 & 49147.67 \\
            \bottomrule
        \end{tabular}%
    }
\end{table*}

To validate the effectiveness of this framework in simplifying the experimental space, we selected a baseline experimental requirement as a test case (see Appendix \ref{appendix-subsec:req_refinement} for details). We conducted comparative experiments using two modes: the experimental design points generated by the Screenwriter Agent versus the Full Factorial Design. The results are presented in \autoref{tab:DoE-result}. The variables $p_1$ to $p_4$ represent the probabilities of the following outcomes: Peace, Peace with Tension, Local War, and Total War, respectively. Data analysis reveals that, compared to the full factorial traversal, the experimental schemes generated by FSTS significantly reduced the experimental scale. Meanwhile, the proportional distribution characteristics of the results maintained high consistency with the full factorial combinations, demonstrating the method's efficiency and reliability.

\subsection{Counterfactual Experiments}
To further validate robustness and adaptability, we introduced counterfactual perturbations that force key events or agent personalities to deviate from historical trajectories and examined the resulting changes in the simulation.

\begin{figure}[H]
    \centering
    \includegraphics[width=1\linewidth]{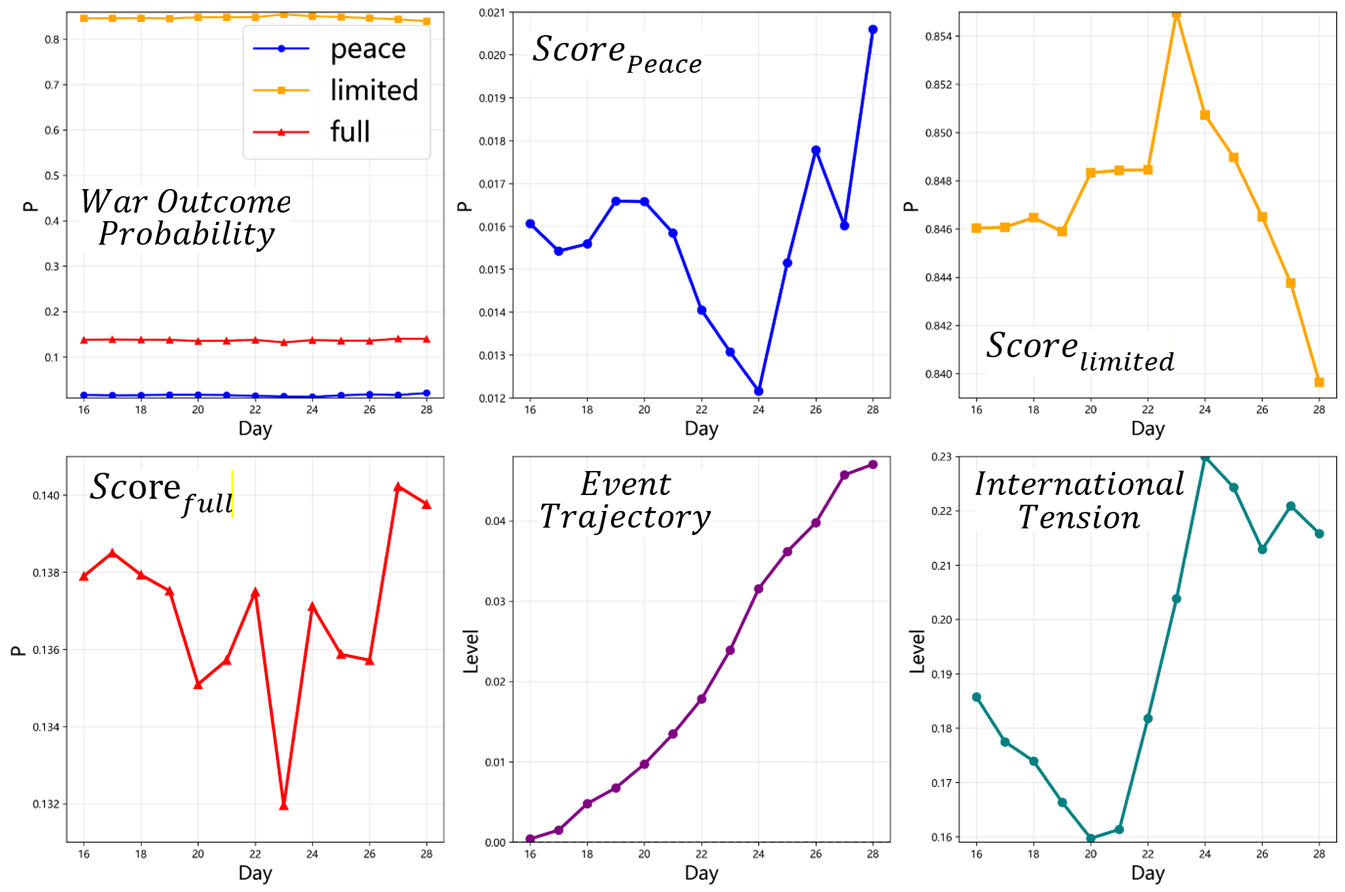}
    \caption{Result of counterfactual experiment.}
    \label{fig:data2}
\end{figure}

Specifically, we constrained ``Kennedy’’ to adopt a consistently hardline stance. The script tracks \textbf{war outcome probabilities}, \textbf{event trajectory}, and \textbf{international tension}, where war outcomes are represented as $[score_{peace}, score_{limited}, score_{full}]$ for peace, limited conflict, and full-scale war. Trends over the 13-day simulation are shown in \autoref{fig:data2}.

Experimental logs indicate that between the 23rd and 24th, small-scale conflicts erupted between the United States and the Soviet Union, leading to a significant deviation in the overall event trajectory. Evaluated via the Sentence-BERT model, the semantic similarity between the Actor Agent's decisions and history was 50.88, while on GPT-5-mini, it was 66.30. Consequently, the final simulation outcome shifted to ``\emph{Limited Conflict: Localized military confrontations occurred but did not escalate into full-scale war.}''

\subsection{Further Analysis}
To assess generalization, we evaluated FSTS on additional scenarios.

\textbf{Digital Service Market Scenario:} We study how the human--machine collaboration ratio affects platform efficiency, cost, and effectiveness under varying environments. Results in \autoref{tab:platform_performance} and \autoref{tab:difficulty_ratio_comparison} suggest that full automation (a high robot ratio) is suboptimal; increasing human nodes yields substantial time-efficiency gains at an acceptable cost increase.

\begin{figure}[t]
    \centering
    \includegraphics[width=1\linewidth]{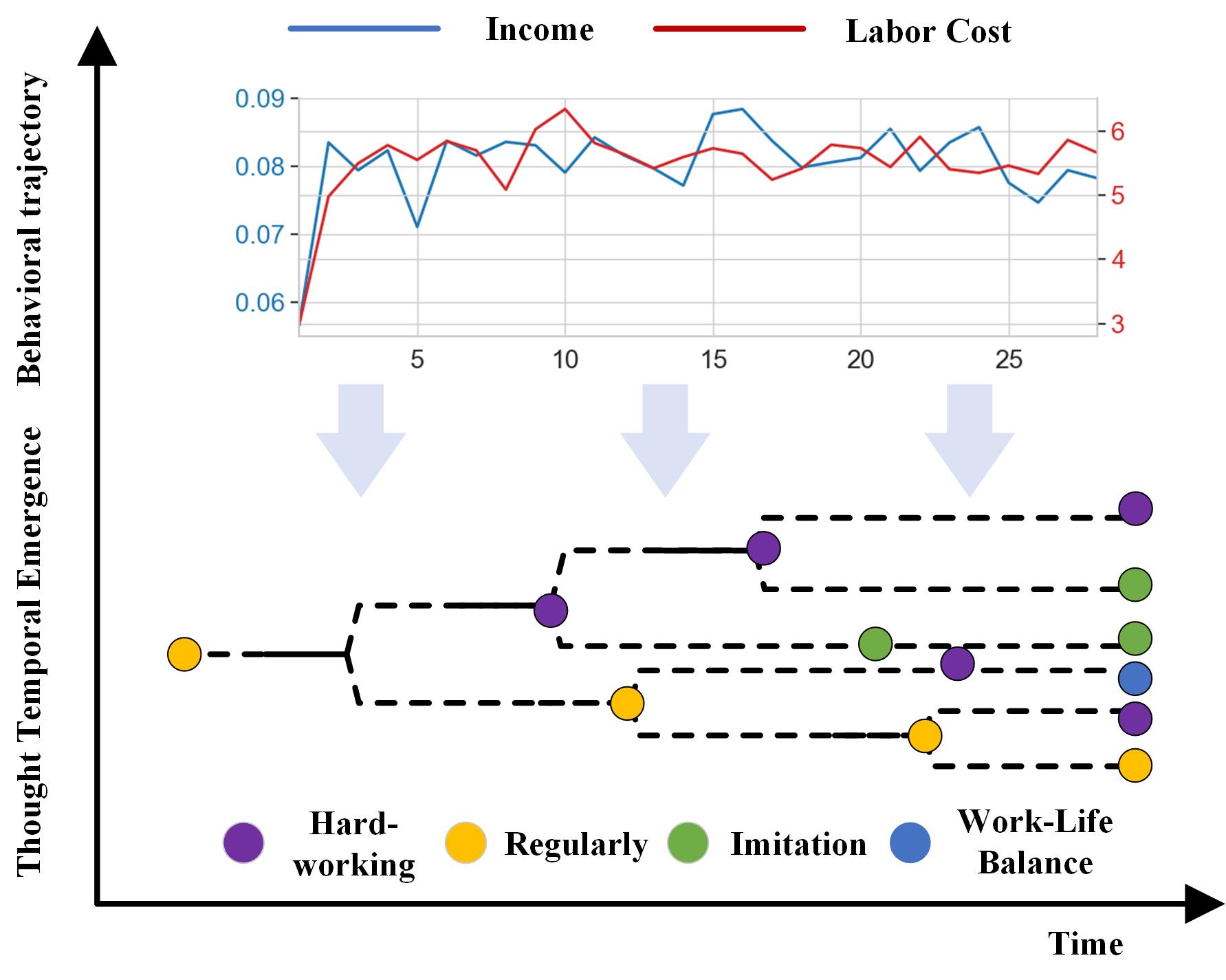}
    \caption{Evolution of Rider Cognition Leading to "Involution".}
    \label{fig:o2o_res}
\end{figure}

\textbf{O2O (Online-to-Offline) Delivery Scenario:} We examine how does the evolution of rider cognition lead to ``Involution''. As shown in \autoref{fig:o2o_res}, the phenomenon stems from a cognitive shift from ``Routine'' to ``Mimicry'' and ``Intensified Effort.'' This escalation drives up labor costs without yielding proportional income gains.

Furthermore, we investigated how role attribute injection affects simulation trajectories, and how the Director Agent's oversight mechanism impacts script generation quality. Detailed experimental setups and results are provided in Appendix \ref{appendix-subsec:role_attribute} and Appendix \ref{appendix-subsec: director_oversight}.

%% file: sections/05_conclusion.tex
\section{Conclusion}
This paper proposes an automated experimental framework based on a \textbf{Screenwriter–Director–Actor} collaborative loop, enabling end-to-end automation from natural-language requirements to experimental design. Multi-scenario case studies demonstrate its validity and generalization for complex social systems, offering a low-barrier and scalable paradigm for AI-driven social science experiments.

%% file: sections/06_limitations.tex
\section{Limitations}
The limitations of this study are mainly reflected in three aspects. First, regarding model ecosystem generalization. While we introduced diverse models (e.g., Qwen, and DeepSeek) during the Script Finalization phase for human alignment evaluation, the core generative pipelines—specifically Script Composition and Actor Generation—still heavily rely on specific GPT-series models. Second, concerning functional boundaries, the existing framework focuses on the automated orchestration of experimental design but has not yet achieved a "full-chain" automated closed loop from user intent analysis to deep result analysis. Third, regarding reasoning robustness, limited by the random nature of LLMs, the system may still experience a decline in logical coherence when dealing with long-range, complex deductions.

%% file: sections/07_ethical_cons.tex
\section{Ethical Considerations}
Our proposed framework for LLM-based social simulation relies solely on publicly available data and does not involve human subjects, ensuring no privacy violations. While we recognize that LLMs may propagate training data biases, we employ structured constraints and multi-agent cross-validation to minimize these effects. This system is a research tool for decision support, not a predictive engine; therefore, its outputs should be interpreted with caution and require human supervision for any real-world application. We advocate for responsible use and provide interpretable outputs to facilitate transparency and prevent the potential simulation of harmful strategies.

%% file: sections/acknowledgments.tex
\section*{Acknowledgments}
This work has been supported in part by National Key Research and Development Program of China (No.2025YFE0216300), National Natural Science Foundation of China (No. 62472306, No. 62441221), Tianjin University's 2024 Special Project on Disciplinary Development (No. XKJS-2024-5-9), Tianjin University Talent Innovation Reward Program for Literature \& Science Graduate Student (C1-2022-010), and Henan Province Key Research and Development Program (No.251111210500), Tianjin University Independent Innovation Project (No.2025XJ3-0043).

%% file: sections/08_appendix.tex
\startcontents[app]  
\addcontentsline{toc}{section}{Appendix Index}
\printcontents[app]{}{1}{\setcounter{tocdepth}{1}}

\section{Supplementary Description}
\subsection{Script Example}
\label{app:script2}

\begin{tcolorbox}[
    colback=gray!5, 
    colframe=gray!60, 
    title=\textbf{Script: Cuban Missile Crisis Simulation},
    arc=2pt, boxrule=0.8pt
]
    \textbf{Research Goal:} By reproducing the Cuban Missile Crisis events, analyze the impact of factors such as national strength, economic situation, technological level, military armament, and leaders' decision-making styles on key war events and final outcomes.
    
    \textbf{Target Type:} Phenomenon Explanation
    
    \textbf{Method:} Scenario analysis (First scenario is historical calibration)
\end{tcolorbox}

\vspace{0.5cm}

\begin{tcolorbox}[
    colback=gray!5, 
    colframe=gray!60, 
    title=\textbf{Input Factors \& Response Variables},
    arc=2pt, boxrule=0.8pt, breakable
]
\begin{itemize}[leftmargin=*, label=$\bullet$]
    \setlength\itemsep{0.2em} 
    
    \item \textbf{Influence Factors:} 
    
    \texttt{us\_conventional\_military\_strength}, 
    \texttt{ussr\_conventional\_military\_strength}, 
    \texttt{us\_strategic\_nuclear\_strength},
    \texttt{ussr\_strategic\_nuclear\_strength}
    \texttt{us\_economic\_capacity}
    \texttt{ussr\_economic\_capacity}
    \texttt{us\_technological\_level}
    \texttt{ussr\_technological\_level}
    \texttt{us\_alliance\_support}
    \texttt{ussr\_alliance\_support}
    \texttt{us\_intelligence\_uncertainty}
    \texttt{ussr\_intelligence\_uncertainty}
    \texttt{us\_deployment\_proximity\_to\_cuba}
    \texttt{ussr\_deployment\_proximity\_to\_cuba}
    \texttt{distance\_us\_ussr}
    \texttt{missile\_deployment\_in\_cuba}
    \texttt{initial\_bilateral\_tension}
    \texttt{perception\_noise\_us}
    \texttt{perception\_noise\_ussr}
    \texttt{us\_leader\_risk\_tolerance}
    \texttt{ussr\_leader\_risk\_tolerance}
    \texttt{us\_leader\_hostility}
    \texttt{ussr\_leader\_hostility}
    \texttt{us\_leader\_transparency}
    \texttt{ussr\_leader\_transparency}
    \texttt{us\_domestic\_political\_pressure}
    \texttt{ussr\_domestic\_political\_pressure}
    \texttt{decision\_temperature\_us}
    \texttt{decision\_temperature\_ussr}

    \item \textbf{Response Variables:} 
    
    \texttt{war\_event\_outcome\_probs}
    \texttt{escalation\_index}
    \texttt{bilateral\_tension\_next}
    \texttt{systemic\_tension\_index}

\end{itemize}
\end{tcolorbox}

\vspace{0.5cm}

\begin{tcolorbox}[
    colback=gray!5, 
    colframe=gray!60, 
    title=\textbf{Parameter Configuration},
    arc=2pt, boxrule=0.8pt, breakable
]

\lstdefinestyle{jsonstyle}{
    basicstyle=\ttfamily\footnotesize,
    breaklines=true, 
    breakatwhitespace=true,
    frame=none,
    columns=fullflexible,
    keepspaces=true,
    literate={,}{,}{1} {]}{]}{1} {[}{[}{1}, 
    moredelim=[is][\bfseries\color{black}]{|}{|} 
}

\begin{lstlisting}[style=jsonstyle]
{
  "us_conventional_military_strength": [0.6, 0.6, 0.5, 0.4, 0.7, 0.65, 0.55, 0.3, 0.8, 0.6, 0.45, 0.6],
  "ussr_conventional_military_strength": [0.8, 0.85, 0.9, 0.75, 0.65, 0.8, 0.95, 0.85, 0.7, 0.6, 0.9, 0.78],
  "us_strategic_nuclear_strength": [0.9, 0.92, 0.88, 0.85, 0.95, 0.9, 0.8, 0.6, 0.97, 0.9, 0.7, 0.85],
  "ussr_strategic_nuclear_strength": [0.85, 0.88, 0.92, 0.8, 0.78, 0.86, 0.95, 0.6, 0.75, 0.7, 0.98, 0.82],
  "us_economic_capacity": [0.9, 0.85, 0.8, 0.75, 0.95, 0.7, 0.6, 0.5, 0.92, 0.88, 0.65, 0.9],
  "ussr_economic_capacity": [0.6, 0.55, 0.5, 0.45, 0.65, 0.7, 0.6, 0.4, 0.72, 0.5, 0.8, 0.58],
  "us_technological_level": [0.9, 0.88, 0.85, 0.8, 0.95, 0.82, 0.7, 0.55, 0.96, 0.9, 0.6, 0.86],
  "ussr_technological_level": [0.7, 0.75, 0.6, 0.55, 0.65, 0.78, 0.85, 0.5, 0.68, 0.72, 0.4, 0.74],
  "us_alliance_support": [0.8, 0.85, 0.7, 0.6, 0.95, 0.5, 0.4, 0.3, 0.9, 0.82, 0.55, 0.75],
  "ussr_alliance_support": [0.4, 0.45, 0.6, 0.5, 0.35, 0.55, 0.7, 0.3, 0.2, 0.6, 0.8, 0.48],
  "us_intelligence_uncertainty": [0.4, 0.3, 0.6, 0.7, 0.2, 0.5, 0.8, 0.25, 0.1, 0.55, 0.65, 0.35],
  "ussr_intelligence_uncertainty": [0.4, 0.35, 0.65, 0.75, 0.3, 0.55, 0.85, 0.4, 0.12, 0.6, 0.7, 0.45],
  "us_deployment_proximity_to_cuba": [0.8, 0.7, 0.2, 0.1, 0.5, 0.6, 0.15, 0.05, 0.9, 0.75, 0.25, 0.82],
  "ussr_deployment_proximity_to_cuba": [0.9, 0.85, 0.95, 0.4, 0.2, 0.75, 0.3, 0.1, 0.85, 0.65, 0.35, 0.88],
  "distance_us_ussr": [0.2, 0.2, 0.25, 0.3, 0.4, 0.1, 0.05, 0.8, 0.5, 0.9, 0.15, 0.6],
  "missile_deployment_in_cuba": [1.0, 0.0, 1.0, 0.0, 0.0, 0.5, 0.0, 0.0, 1.0, 0.0, 1.0, 1.0],
  "initial_bilateral_tension": [0.5, 0.2, 0.7, 0.6, 0.4, 0.65, 0.3, 0.15, 0.55, 0.45, 0.8, 0.6],
  "perception_noise_us": [0.1, 0.05, 0.2, 0.15, 0.05, 0.6, 0.4, 0.02, 0.12, 0.5, 0.3, 0.08],
  "perception_noise_ussr": [0.1, 0.05, 0.25, 0.15, 0.2, 0.65, 0.45, 0.03, 0.14, 0.55, 0.35, 0.1],
  "us_leader_risk_tolerance": [0.5, 0.4, 0.6, 0.55, 0.65, 0.7, 0.3, 0.2, 0.8, 0.5, 0.45, 0.58],
  "ussr_leader_risk_tolerance": [0.6, 0.4, 0.8, 0.7, 0.4, 0.75, 0.35, 0.25, 0.85, 0.6, 0.5, 0.65],
  "us_leader_hostility": [0.3, 0.2, 0.4, 0.35, 0.25, 0.6, 0.5, 0.1, 0.45, 0.28, 0.7, 0.32],
  "ussr_leader_hostility": [0.5, 0.2, 0.8, 0.6, 0.45, 0.65, 0.55, 0.15, 0.6, 0.35, 0.85, 0.48],
  "us_leader_transparency": [0.6, 0.8, 0.4, 0.5, 0.8, 0.3, 0.25, 0.7, 0.55, 0.45, 0.2, 0.65],
  "ussr_leader_transparency": [0.4, 0.7, 0.3, 0.45, 0.5, 0.3, 0.2, 0.65, 0.4, 0.5, 0.15, 0.62],
  "us_domestic_political_pressure": [0.6, 0.3, 0.6, 0.5, 0.4, 0.8, 0.2, 0.1, 0.65, 0.55, 0.9, 0.58],
  "ussr_domestic_political_pressure": [0.7, 0.3, 0.9, 0.7, 0.6, 0.8, 0.4, 0.25, 0.75, 0.5, 0.95, 0.68],
  "decision_temperature_us": [0.3, 0.5, 0.35, 0.4, 0.3, 0.8, 0.2, 0.6, 0.25, 0.45, 0.9, 0.38],
  "decision_temperature_ussr": [0.4, 0.5, 0.25, 0.3, 0.4, 0.9, 0.22, 0.55, 0.3, 0.48, 0.85, 0.42]
}
\end{lstlisting}

\end{tcolorbox}

\begin{table*}[t]
    \centering
    \caption{Comparison of Different Agent Frameworks}
    \label{tab:framework_comparison}
    \resizebox{\textwidth}{!}{%
        \begin{tabular}{@{}lp{5cm}ccp{5cm}@{}}
            \toprule
            \textbf{Framework} & \textbf{Agent Generation Mechanism} & \textbf{Agents} & \textbf{Self-Correction} & \textbf{Limitations} \\ 
            \midrule
            AutoGPT & Think-Act-Feedback Loop & 1 & $\checkmark$ & Poor stability in task execution \\
            Camel & Role assignment based on scenarios & 2 & $\times$ & High dialogue openness; prone to deviating from task goals \\
            ChatDev & Strictly predefined multi-agent collaboration & 6 & $\checkmark$ & Limited scenarios; applicable only to software development \\
            MetaGPT & Agent generation based on role templates & Unlimited & $\times$ & Over-engineered workflows; limited generalizability \\
            AutoGen & User-defined configuration & Unlimited & $\times$ & Cumbersome configuration; high dependency on parameters \\
            LangGraph & Code-based agent customization & Unlimited & $\checkmark$ & Requires user programming proficiency \\
            FSTS (Ours) & Dynamic generation based on scripts \& experimental scenarios & Unlimited & $\checkmark$ & Relies on high-quality scripts \\ \bottomrule
        \end{tabular}%
    }
\end{table*}

\begin{table*}[t]
    \centering
    \caption{Representative events of the Cuban Missile Crisis}
    \label{tab:cuban_crisis_timeline_en}

    \begin{tabularx}{\textwidth}{lXX}
        \toprule
        \textbf{Date} & \textbf{Event} & \textbf{Action taken by Country/Leader} \\
        \midrule
        Oct. 16 & U.S. U-2 reconnaissance aircraft detects Soviet missile deployment in Cuba. & Established ExComm to institute a top-secret decision-making framework. \\
        Oct. 18 & Soviet Foreign Minister Gromyko visits the U.S. and denies missile deployment. & JFK withheld intelligence and feigned ignorance to probe Soviet intentions. \\
        Oct. 20 & The U.S. formulates a response plan. & Adopted ExComm's recommendation; authorized a ``Naval Quarantine'' over air strikes. \\
        Oct. 22 & Kennedy publicly reveals Soviet missile deployment in Cuba. & Delivered televised address to inform the nation and declare resolve for the blockade. \\
        Oct. 23 & U.S. prepares to implement the quarantine. & Secured OAS endorsement for legal backing; fully deployed the military blockade. \\
        Oct. 24 & U.S. blockade goes into effect. & Soviet ships turned back at the quarantine line; military standoff ensued. \\
        Oct. 25 & Confrontation at the UN. & US unveiled aerial photographic evidence at the UN to dominate the diplomatic narrative. \\
        Oct. 26 & The Soviet side signals goodwill to the U.S. & Khrushchev proposed a ``non-invasion pledge for missile removal'' deal. \\
        Oct. 27 & A U.S. reconnaissance plane is shot down over Cuba; military advises retaliation. & JFK ignored the letter, responding only to the initial conciliatory proposal. \\
        Oct. 28 & Khrushchev announces agreement to withdraw missiles from Cuba via broadcast. & Concluded a secret deal to resolve the crisis. \\
        \bottomrule
    \end{tabularx}
\end{table*}

\subsection{Human Alignment Study for LLM-as-a-Judge}
\label{subsec:human_alignment_study}
We randomly selected 50 sets of scripts generated by the Screenwriter Agent prior to the final script finalization phase in FSTS. Each set contained 10 candidate scripts. Via questionnaire surveys, we invited 25 sociologists and 25 computer scientists to finalize the script for each set. Concurrently, we employed GPT-5 mini, Qwen3-235B, DeepSeek-V3.2, Gemini 3 Pro, and GPT-5.1 as Director Agents to independently select the experimental scripts.

We utilized Krippendorff's $\alpha$ to evaluate the agreement among human scientists, among AI models, and between the two groups. As shown in \autoref{tab:krippendorff_alpha}, the experimental scripts selected by the FSTS Director Agent exhibit a high degree of consistency with the choices made by human scientists.

\begin{table*}[t]
    \centering
    \caption{Krippendorff's $\alpha$ Across Different Evaluator Groups.}
    \label{tab:krippendorff_alpha}
    \small
    \begin{tabularx}{\textwidth}{
        >{\hsize=1\hsize\centering\arraybackslash}X 
        >{\hsize=0.9\hsize\centering\arraybackslash}X 
        >{\hsize=1.4\hsize\centering\arraybackslash}X 
        >{\hsize=0.9\hsize\centering\arraybackslash}X 
        >{\hsize=0.8\hsize\centering\arraybackslash}X
    }
    \toprule
        \textbf{Computer Scientists} & \textbf{Social Scientists} & \textbf{FSTS Director Agent} & \textbf{Human Scientists} & \textbf{Overall}\\
    \midrule
        0.8891 & 0.8849 & 0.9783 & 0.7665 & 0.7711\\
    \bottomrule
    \end{tabularx}
\end{table*}

\section{Additional Results}
\label{appendix-sec:additional_result}
\subsection{FSTS-Guided Requirement Refinement}
\label{appendix-subsec:req_refinement}

\begin{figure}[H]
    \centering
    \includegraphics[width=1\linewidth]{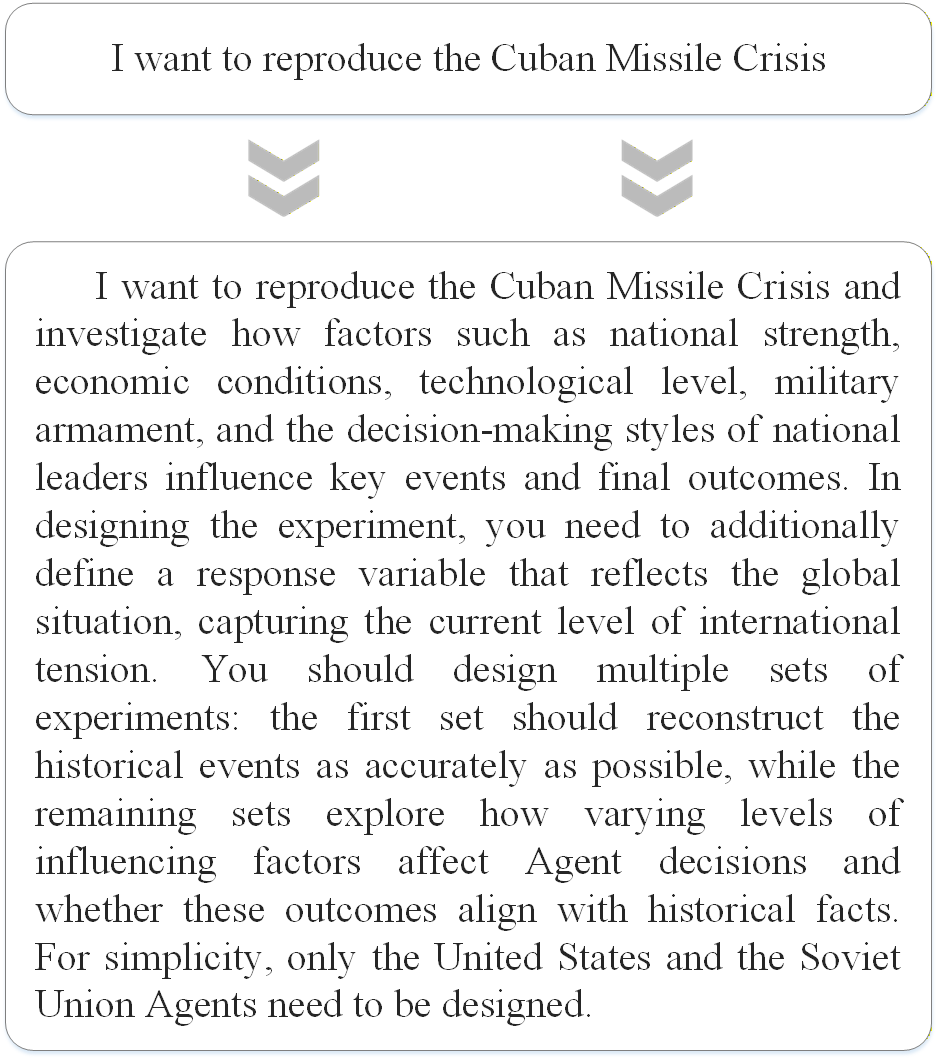}
    \caption{Example of requirement refinement.}
    \label{fig:req_change}
\end{figure}

\subsection{Q\&A on the digital services market scenario}
\begin{figure}[H]
    \centering
    \includegraphics[width=1\linewidth]{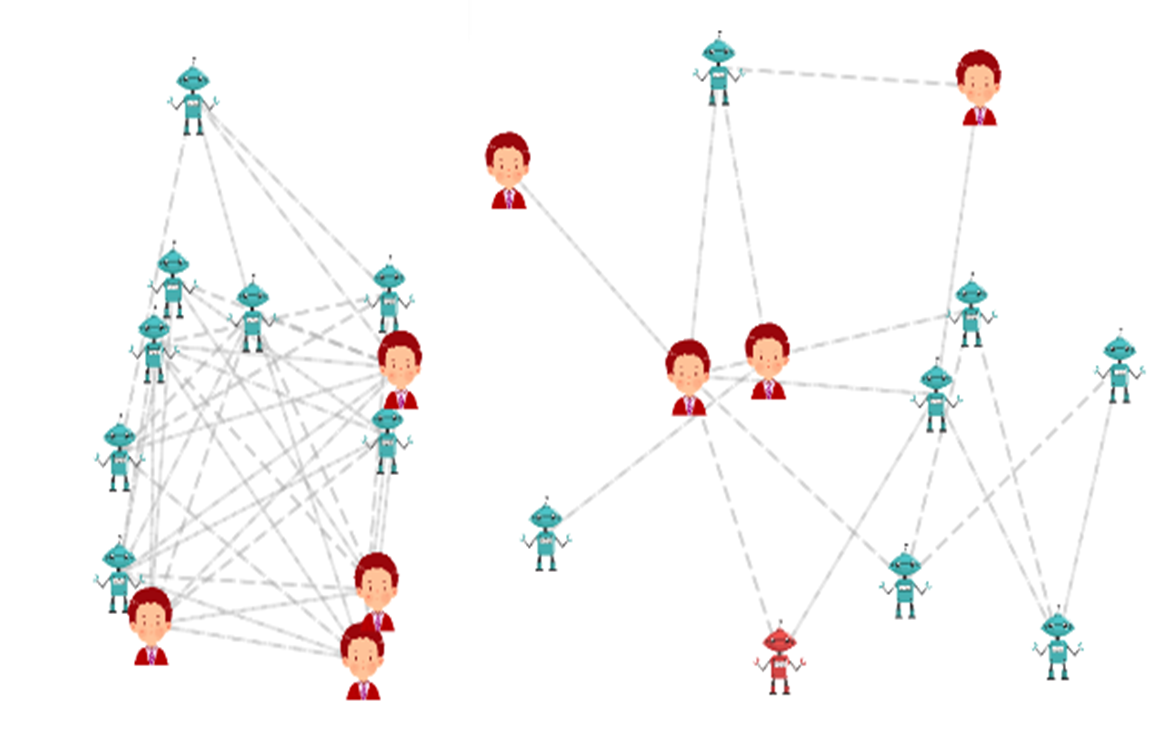}
    \caption{Network conditions during system congestion.}
    \label{fig:network}
\end{figure}
\textbf{Question:} To investigate the mechanism by which the human-machine collaboration ratio impacts the platform's overall operational efficiency, cost, and effectiveness under different external environments.

\noindent\textbf{Answer:} As shown in \autoref{tab:platform_performance} and \autoref{tab:difficulty_ratio_comparison}, the system resource structure is clearly unbalanced in scenarios involving high-complexity orders. \autoref{fig:network} reveals that human clerks were in a chronic state of high load, whereas robot agents remained largely idle. This dependency resulted in a substantial backlog of orders at the human processing stage and a waste of computational resources. A comparison of experimental results with different human-machine ratios demonstrates that appropriately increasing the proportion of human clerks significantly mitigates these bottlenecks. Therefore, sole reliance on automation (a high robot ratio) is not the optimal solution. Optimizing collaboration by increasing the number of human nodes is an effective strategy to trade an acceptable cost increment for significant gains in time efficiency.

\subsection{Q\&A on the O2O delivery scenario}
\textbf{Question:} To investigate how the collective mental state of delivery riders evolves over time with changes in labor expenditure and income, and how this ultimately leads to the phenomenon of ``Involution.''

\noindent\textbf{Answer:} In the \autoref{fig:o2o_res}, different colors denote riders in distinct cognitive or operational states:

\begin{itemize}
    \item \textbf{Yellow circles}: Represent the initial state, i.e., riders working ``Regularly'' (Routine).
    \item \textbf{Purple circles}: Represent riders entering a ``Hard-working'' state.
    \item \textbf{Green circles}: Represent riders engaging in ``Imitation'' of others.
    \item \textbf{Blue circles}: Represent riders maintaining a ``Work-Life Balance.''
    \item \textbf{Dashed lines}: Indicate the interactions and transition processes between different states.
\end{itemize}
Initially, riders predominantly occupy the ``Regularly'' state (yellow nodes). As time progresses, interactions and mutual influences occur among riders. The graph illustrates a progressive shift where an increasing number of riders transition from the initial state to ``Imitation'' (green) or ``Hard-working'' (purple) states. Over time, while the riders' Labor Cost rises, their Income stabilizes (plateaus) after a few days. Although the majority of riders are influenced to intensify their efforts or mimic high-intensity workers—leading to an increase in overall working hours—their actual income does not rise proportionally due to the finite total volume of orders. Only a minority (blue) maintain their original state. This phenomenon of ``expending more labor (increased travel distance) without a proportional increase in income'' characterizes the essence of ``Involution'' as revealed through cognitive analysis.

\subsection{Quantitative Analysis of the Impact of "Role Attribute Injection" on Simulation Trajectories}
\label{appendix-subsec:role_attribute}
\textbf{Experimental Setup:} We selected the Cuban Missile Crisis and O2O Delivery scenarios, utilizing GPT-5 mini as the evaluation backbone, to compare the performance of the following three configurations across 30 repeated experiments:

\begin{itemize}
    \item \textbf{Group A (Full Attribute Injection):} Includes domain knowledge injection, emotional mapping (bounded rationality), adaptive goal planning, and heterogeneous cognitive customization (attribute orthogonalization).
    \item \textbf{Group B (Partial Injection - No Emotion/Domain):} Removes emotional mapping and domain knowledge; agents rely solely on the model's general common sense and basic role settings.
    \item \textbf{Group C (Zero Injection - Baseline):} Removes all injection mechanisms, providing agents only with simple identity labels to simulate generic multi-turn LLM dialogues.
\end{itemize}

\begin{table*}[htbp]

    \centering
    \caption{The Impact of the Director Agent's Oversight Mechanism on Script Quality.}
    \label{appendix-tab:director_mechanism}
    \small
    
    \renewcommand{\arraystretch}{1.3}
    
    \renewcommand{\tabularxcolumn}[1]{m{#1}}
    
    \begin{tabularx}{\textwidth}{
        >{\hsize=1.2\hsize\raggedright\arraybackslash}X 
        >{\hsize=0.9\hsize\centering\arraybackslash}X 
        >{\hsize=1.1\hsize\centering\arraybackslash}X 
        >{\hsize=0.6\hsize\centering\arraybackslash}X 
        >{\hsize=0.8\hsize\centering\arraybackslash}X 
        >{\hsize=1.4\hsize\centering\arraybackslash}X
    }
    \toprule
        \textbf{Experimental Setup} & \textbf{Script Generation Time (min)} & \textbf{Proportion of High-Quality Scripts} & \textbf{Result Entropy} & \textbf{Actor Attribute Alignment} & \textbf{Historical Trajectory Semantic Similarity}\\
    \midrule
        With Director Oversight & 102.39 & 96.7\% & 0.72 & 0.88 & 72.50\\
        Without Director Oversight & 5.32 & 43.3\% & 1.68 & 0.46 & 51.40\\
        Without Feedback Mechanism & 12.21 & 63.3\% & 1.03 & 0.62 & 59.60\\
    \bottomrule
    \end{tabularx}
\end{table*}

\textbf{Experimental Results:}

\begin{itemize}
    \item \textbf{In the Cuban Missile Crisis scenario:} Group A proved to be the core element in ensuring the authenticity of the simulation trajectory. Within this group, the agents' decisions at critical time nodes exhibited an extremely high degree of alignment with historical trajectories, achieving a GPT-5 mini semantic similarity score of 72.5. In contrast, Group B, which lacked emotional mapping and professional background, saw its score drop to 63.2, while Group C, equipped only with simple identity labels, plummeted to 42.1. The data demonstrates that agents lacking state and emotional mapping are unable to simulate the cognitive biases of real humans under pressure. Their behavioral trajectories lean toward generic linguistic alignment rather than authentic social gaming.
    
    \item \textbf{In the O2O Delivery scenario:} At the micro-level for Group A, attribute injection endowed rider agents with bounded rationality and the right of refusal. Agents were able to make game-theoretic choices regarding dispatches based on current load, income expectations, and physiological states, successfully simulating the non-linear characteristic of ``increased labor expenditure with plateauing income.'' Conversely, when facing extreme high-pressure order flows, Groups B and C---lacking cognitive architectural constraints---tended to either blindly accept all orders or reject them without reason. Their decision-making rationality significantly deteriorated, leading to frequent system capacity collapses.
\end{itemize}

\subsection{The Impact of the Director Agent's Oversight Mechanism on Script Quality}
\label{appendix-subsec: director_oversight}

\textbf{Experimental Setup:} We consistently used GPT-5 mini and GPT-4o as the backbones for the Director Agent and Screenwriter Agent, respectively. We investigated the impact on script generation under three conditions: with Director Agent oversight, without Director Agent oversight, and without a feedback mechanism. Using the user requirements mentioned in the text for the Cuban Missile Crisis scenario, we conducted 30 repeated experiments for each of the three conditions.

As comparative considerations, we evaluated the quality of the generated scripts, the actor generation process, and the actors' performances on stage. The resulting data is presented in the \autoref{appendix-tab:director_mechanism}.

\vspace{1em}

\noindent\textbf{Metric Definitions:}

\begin{itemize}
    \item \textbf{Script Generation Time:} The average time taken by the Screenwriter Agent to generate a script, measured in minutes.
    \item \textbf{Proportion of High-Quality Scripts:} The percentage of scripts that have a correct JSON format and contain the complete $\langle G, I, R, D, L \rangle$ 5-tuple.
    \item \textbf{Result Entropy:} Represents the stability and consistency of the influencing factors and response variables within the generated scripts. The calculation formula is $H_{res} = - \sum_{i=1}^{n} P(x_i) \log_2 P(x_i)$, where $x_i$ represents the $i$-th specific script configuration (i.e., a specific combination of influencing factor sets and response variable sets).
    \item \textbf{Actor Attribute Alignment:} The degree to which the Agent attributes generated by the Actor Factory align with the requirements of the ``Storyline ($L$)'' and ``Design Points ($D$)'' in the script. This metric is evaluated by GPT-5 mini and is a value between 0 and 1, with higher values indicating better alignment. Prior to the evaluation, we manually provided two scripts with alignment scores of 0.3 and 0.8 as baseline references for the LLM.
    \item \textbf{Historical Trajectory Semantic Similarity:} We used GPT-5 mini to calculate the alignment between the simulated actions and the historical records (Table 6).
\end{itemize}

\section{Scenario Description}
\subsection{Cuban Missile Crisis Simulation Scenario}
This study selects the Cuban Missile Crisis as the core experimental scenario. This event represents not only a high-stakes geopolitical confrontation but also a Complex Social Adaptive System that poses significant modeling challenges. As a historical event, it offers distinct advantages, including rich data availability, multi-layered modeling potential, a clear timeline with critical decision nodes, and traceable outcomes.

The complexity of the modeling is primarily reflected in the following three dimensions:

\textbf{Cognitive Heterogeneity and Dynamics.} Unlike particle simulations characterized by homogeneity, every Agent in this scenario possesses an independent cognitive architecture. Agents perceive external information through a ``Global Channel'' while being simultaneously driven by their intrinsic political stances and psychological states. This interaction between endogenous thinking and the exogenous environment renders the decision-making process full of irrationality and uncertainty.

\textbf{Interaction Sparsity and Temporal Sensitivity.} The experiment uses ``days'' as discrete time steps to simulate the high-tension atmosphere where every second counts. Information transmission occurs through restricted channels, compelling Agents to make decisions within a ``fog of incomplete information.'' Any minor misjudgment can be amplified into a nuclear war through non-linear feedback mechanisms.

\textbf{Quantifiable Representation of Macro-Emergence.} To capture the impact of micro-interactions on the macro situation, the system introduces the ``International Tension Index'' as a global state variable. This index is not a simple linear weighting; rather, it is dynamically calculated based on daily Agent interaction results, reflecting in real-time the system's distance from the ``brink of collapse.''

\subsection{O2O Delivery Scenario}
This study selects O2O instant delivery services as the core experimental scenario. This domain represents not merely a large-scale logistics scheduling problem but a Complex Adaptive System (CAS) presenting significant modeling challenges. Within this scenario, the experimental platform constructs an open environment that supports user-defined distributions of merchants, users, and riders, simulating the highly dynamic supply-demand matching network of the real world.

The complexity of the modeling is primarily reflected in the following three dimensions:

Agent Heterogeneity and Autonomous Decision-Making: Unlike the homogeneous particles in traditional path planning, Rider Agents in this scenario possess independent cognitive architectures and attribute differences. The system utilizes an ``ExtraAttrSetter'' to endow riders with multi-dimensional heterogeneous attributes, including gender, age, maximum order capacity, and personalized descriptions. Riders are no longer mechanical executors of platform commands but active strategic agents with the Right of Refusal. When facing a platform dispatch, riders evaluate the order based on their current state (location, existing order load) and internal logic. They may make a decision to reject the order and provide specific reasons. This micro-level autonomous gaming increases the uncertainty of system scheduling.

Environmental Stochasticity and Temporal Sensitivity: The experiment uses discrete Time Steps as the unit to simulate the time-critical pressure of instant delivery. The generation of order flows is not linear but follows a stochastic distribution model consistent with reality, integrating multiple variables such as Peak periods, Weekend effects, and Area Tiers. Agents must interact in real-time within this highly non-stationary, tidal environment, where any delay or rejection in a single link can trigger a butterfly effect, impacting subsequent capacity allocation.

Holistic Observability of Macro-States: To capture the emergent impact of micro-interactions on macro-efficiency, the system establishes a comprehensive state tracking mechanism. The platform functions as a ``Panopticon,'' calculating and recording the spatial trajectories, order loads, and income of all riders in real-time at every time step. This data flow not only reflects the system's real-time load but also quantifies the system's elasticity and service boundaries under pressure through the closed-loop logic of ``Dispatch—Feedback—Redispatch.''

\subsection{Digital Service Market Scenario}
The Data Service Market represents not only an efficient platform for digital government collaboration but also a typical Multi-Agent Complex Social System. In this scenario, the experimental platform constructs a three-tier governance environment consisting of ``Platform—Agency—Agent,'' simulating the cross-departmental and cross-hierarchical data service collaboration network in the real world.

The complexity of the modeling is primarily reflected in the following three dimensions:

\textbf{Multi-level Collaboration and Micro-Heterogeneity.} Transcending single-layer, flat multi-agent systems, this scenario builds a hierarchical organizational architecture. Externally, the system faces continuous dynamic demands (orders); internally, it nests a three-layer structure: the Top-level Platform is responsible for macro-scheduling and task allocation; Middle-level Agencies (8 heterogeneous institutions) specialize in specific types of government services and possess independent internal network topologies (e.g., Small-World or Scale-Free networks); Bottom-level Agents (clerks and robots) handle specific tasks based on heterogeneous attributes (e.g., efficiency, maintenance costs, human-machine ratio). This nested modeling requires the system to simultaneously handle macro-level inter-agency collaboration and micro-level individual behavioral gaming.

\textbf{Full-Process Automated Experimental Closed-Loop.} The platform achieves a full-process closed loop from Crowd Intelligence Modeling to intervention and control. It supports user-defined environmental parameters (e.g., order arrival rates following Queuing Theory distributions) and agent attributes, while integrating visual construction tools for Structural Equation Modeling (SEM). The experiment is not merely a running process but a comprehensive scientific workflow containing ``Environment Design—Agent Design—Experiment Design—System Execution—Intervention Control—Analysis—Optimization—Reporting.'' Notably, it introduces a ``Fishbone-style'' interface for experimental design, supporting visual causal modeling of goals, influencing factors, and response variables, which greatly enhances the scientific rigor and interpretability of the design.

\textbf{Dynamic Intervention and Anomaly Detection Mechanisms.} To ensure system robustness, the platform incorporates a high-precision anomaly detection module. By integrating a dual mechanism for system-level and individual-level detection, the system can monitor fluctuations in key metrics such as service efficiency, Value Entropy, and productivity in real-time. The system supports Online Intervention, allowing users to dynamically adjust parameters or inject sudden events (e.g., adjusting order strategies or wage systems) during the simulation.

\section{Prompt Templates for FSTS Agents}

\subsubsection{Screenwriter}

\begin{tcolorbox}[colback=gray!10,colframe=gray!60,boxrule=0.6pt,arc=2pt,breakable]

[\textbf{System Prompt}]\\
You are conducting a requirements analysis for the deduction of complex social model systems. You are to act as a Senior Requirements Engineer. While you may draw upon your own experience, you must strictly prioritize content relevant to the simulation process. During this process, you may encounter references to ``experimental methods''; please note that these refer exclusively to computational experiment methods.

\vspace{0.3em}
[\textbf{User Prompt}]\\
- You are required to screen and evaluate based on the currently proposed experimental request, the objectives of the request, the influencing factors, the response variables, and the experimental analysis methods. Your generated scheme should focus on <focus>.\\
The user's request is: <req>\\
- When responding, you must adopt the following JSON format. Do not provide explanations for any variables.

\vspace{0.3em}
\noindent
\textbf{goal:} The objective of the user's current experiment or hypothesis...\\
\textbf{influence\_factor:} Influencing factors...\\
\textbf{response\_var:} Response variables...\\
\textbf{formula:} The corresponding formulas between influencing factors and response variables...\\
\textbf{exp\_params:} The format for experimental parameters is JSON...\\
\textbf{story\_line:} The main storyline...
\end{tcolorbox}

\subsubsection{Director}
\begin{tcolorbox}[colback=gray!10,colframe=gray!60,boxrule=0.6pt,arc=2pt,breakable]
[\textbf{System Prompt}]

You are currently conducting requirements analysis and Agent design for the simulation of complex social model systems. Prior to this, I have proposed a reasonable experimental scheme based on user requirements and generated the Agents required for the experiment according to both the requirements and this scheme.

Your task is: Referencing the provided `Requirements' and `Experimental Scheme', judge whether the quantity and attributes of the generated Agents are reasonable.

If they are unreasonable, please provide the reasons using the most concise language possible.

[\textbf{User Prompt}]

User Requirements: <req>

Generated Experimental Design Scheme: <exp\_plan>

Generated Agents: <agent\_list>

- Your response must be in JSON format and cover two scenarios:

Case 1: If you deem the agent scheme I generated to be reasonable, the response format is: `is\_reasonable'': 1 ``reason'': The justification for its rationality. reason must be a string in Chinese.

Case 2: If you deem the previous analysis unreasonable, the response format is as follows: ``is\_reasonable'': 0 ``reason'': The reason why the agent team is unreasonable. Please be as concise as possible and provide clear modification suggestions. reason must be a string in Chinese.
\end{tcolorbox}

\section{Hyperparameter settings}
\label{appendix-sec:hyperparameter}
We configured the hyperparameters for both GPT-4o and GPT-5 Mini to ensure a fair comparison. The temperature was set to $0.7$ to allow for diverse generation, while Top-p was kept at $1.0$. To mitigate repetitive loops, we applied a Frequency Penalty of $0.0$ (unless otherwise noted). For reproducibility, we fixed the random seed to $42$ across all API calls. The maximum generation length was restricted to $4,096$ tokens.

%% file: custom.bib
@article{tran2025multi,
  title={Multi-agent collaboration mechanisms: A survey of llms},
  author={Tran, Khanh-Tung and Dao, Dung and Nguyen, Minh-Duong and Pham, Quoc-Viet and O'Sullivan, Barry and Nguyen, Hoang D},
  journal={arXiv preprint arXiv:2501.06322},
  year={2025}
}

@article{xu2024ai,
  title={AI for social science and social science of AI: A survey},
  author={Xu, Ruoxi and Sun, Yingfei and Ren, Mengjie and Guo, Shiguang and Pan, Ruotong and Lin, Hongyu and Sun, Le and Han, Xianpei},
  journal={Information Processing \& Management},
  volume={61},
  number={3},
  pages={103665},
  year={2024},
  publisher={Elsevier}
}

@article{xue2021computational,	
  title={Computational experiments for complex social systems—Part I: The customization of computational model},
  author={Xue, Xiao and Chen, Fangyi and Zhou, Deyu and Wang, Xiao and Lu, Min and Wang, Fei-Yue},
  journal={IEEE Transactions on Computational Social Systems},
  volume={9},
  number={5},
  pages={1330--1344},
  year={2021},
  publisher={IEEE}
}

@article{xue2024computational,
  title={Computational experiments for complex social systems: Experiment design and generative explanation},
  author={Xue, Xiao and Zhou, Deyu and Yu, Xiangning and Wang, Gang and Li, Juanjuan and Xie, Xia and Cui, Lizhen and Wang, Fei-Yue},
  journal={IEEE/CAA Journal of Automatica Sinica},
  volume={11},
  number={4},
  pages={1022--1038},
  year={2024},
  publisher={IEEE}
}

@article{wolf2023decision,
  title={The Decision Theatre Triangle for societal challenges—An example case and research needs},
  author={Wolf, Sarah and F{\"u}rst, Steffen and Geiges, Andreas and Laublichler, Manfred and Mielke, Jahel and Steudle, Gesine and Winter, Konstantin and Jaeger, Carlo},
  journal={Journal of Cleaner Production},
  volume={394},
  pages={136299},
  year={2023},
  publisher={Elsevier}
}

@inproceedings{ qin2023chatgpt,
Author = {Qin, Chengwei and Zhang, Aston and Zhang, Zhuosheng and Chen, Jiaao and
   Yasunaga, Michihiro and Yang, Diyi},
Editor = {Bouamor, H and Pino, J and Bali, K},
Title = {Is ChatGPT a General-Purpose Natural Language Processing Task Solver?},
Booktitle = {2023 CONFERENCE ON EMPIRICAL METHODS IN NATURAL LANGUAGE PROCESSING,
   EMNLP 2023},
Year = {2023},
Pages = {1339-1384},
Note = {Conference on Empirical Methods in Natural Language Processing (EMNLP),
   Singapore, SINGAPORE, DEC 06-10, 2023},
Organization = {Apple; Colossal AI; Google Res; GTCOM; King Salman Global Acad Arabic
   Language; LivePerson; SONY; Ahrefs; Alibaba Cloud; Amazon Sci; Baidu;
   ByteDance; Cohere; Megagon Labs; NEC; ANT Grp; Bloomberg Engn; HUAWEI; J
   P Morgan Chase \& Co; Salesforce; SAP; AiXplain; Duolingo; Jenni;
   Translated; Adobe; Babelscape; ModelBest; Nyonic; Mercari},
ISBN = {979-8-89176-060-8},
ResearcherID-Numbers = {Yasunaga, Michihiro/GPW-9499-2022
   Zhang, Zhuosheng/AAF-4919-2020
   qin, chengwei/MBV-9309-2025},
ORCID-Numbers = {Zhang, Zhuosheng/0000-0002-4183-3645
   },
Unique-ID = {WOS:001275019901029},
}

@inproceedings{park2023generative,
  title={Generative agents: Interactive simulacra of human behavior},
  author={Park, Joon Sung and O'Brien, Joseph and Cai, Carrie Jun and Morris, Meredith Ringel and Liang, Percy and Bernstein, Michael S},
  booktitle={Proceedings of the 36th annual acm symposium on user interface software and technology},
  pages={1--22},
  year={2023}
}

@article{liu2024llms,
  title={Are LLMs good at structured outputs? A benchmark for evaluating structured output capabilities in LLMs},
  author={Liu, Yu and Li, Duantengchuan and Wang, Kaili and Xiong, Zhuoran and Shi, Fobo and Wang, Jian and Li, Bing and Hang, Bo},
  journal={Information Processing \& Management},
  volume={61},
  number={5},
  pages={103809},
  year={2024},
  publisher={Elsevier}
}

@inproceedings{liu2024we,
  title={" we need structured output": Towards user-centered constraints on large language model output},
  author={Liu, Michael Xieyang and Liu, Frederick and Fiannaca, Alexander J and Koo, Terry and Dixon, Lucas and Terry, Michael and Cai, Carrie J},
  booktitle={Extended Abstracts of the CHI Conference on Human Factors in Computing Systems},
  pages={1--9},
  year={2024}
}

@inproceedings{casalicchio2024ai,
  title={AI-CRAS: AI-driven Cloud Service Requirement Analysis and Specification},
  author={Casalicchio, Emiliano and Cotumaccio, Alberto},
  booktitle={2024 IEEE International Conference on Cloud Engineering (IC2E)},
  pages={11--21},
  year={2024},
  organization={IEEE}
}

@inproceedings{santos2024we,
  title={Are we testing or being tested? exploring the practical applications of large language models in software testing},
  author={Santos, Robson and Santos, Italo and Magalhaes, Cleyton and de Souza Santos, Ronnie},
  booktitle={2024 IEEE Conference on Software Testing, Verification and Validation (ICST)},
  pages={353--360},
  year={2024},
  organization={IEEE}
}

@inproceedings{ chen2023autoagents,
Author = {Chen, Guangyao and Dong, Siwei and Shu, Yu and Zhang, Ge and Sesay,
   Jaward and Karlsson, Boerje and Fu, Jie and Shi, Yemin},
Editor = {Larson, K},
Title = {AutoAgents: A Framework for Automatic Agent Generation},
Booktitle = {PROCEEDINGS OF THE THIRTY-THIRD INTERNATIONAL JOINT CONFERENCE ON
   ARTIFICIAL INTELLIGENCE, IJCAI 2024},
Year = {2024},
Pages = {22-30},
Note = {33rd International Joint Conference on Artificial Intelligence (IJCAI),
   Jeju, SOUTH KOREA, AUG 03-09, 2024},
Organization = {Int Joint Conf Artifical Intelligence},
ISBN = {978-1-956792-04-1},
ResearcherID-Numbers = {Dong, Siwei/AFX-5505-2022
   Chen, Guangyao/KOD-2994-2024},
Unique-ID = {WOS:001347142800003},
}

@inproceedings{hong2024metagpt,
  title={MetaGPT: Meta programming for a multi-agent collaborative framework},
  author={Hong, Sirui and Zhuge, Mingchen and Chen, Jonathan and Zheng, Xiawu and Cheng, Yuheng and Zhang, Ceyao and Wang, Jinlin and Wang, Zili and Yau, Steven Ka Shing and Lin, Zijuan and others},
  year={2024},
  organization={International Conference on Learning Representations, ICLR}
}

@article{yang2023auto,
  title={Auto-gpt for online decision making: Benchmarks and additional opinions},
  author={Yang, Hui and Yue, Sifu and He, Yunzhong},
  journal={arXiv preprint arXiv:2306.02224},
  year={2023}
}

@article{wang2004artificial,
  title={Artificial societies, computational experiments, and parallel systems a discussion on computational theory of complex social-economic systems},
  author={Wang, Fei-Yue},
  journal={Fuza Xitong yu Fuzaxing Kexue(Complex Systems and Complexity Science)},
  volume={1},
  number={4},
  pages={25--35},
  year={2004}
}

@article{wang2004computational,
  title={Computational experiments for behavior analysis and decision evaluation of complex systems},
  author={Wang, Fei-Yue},
  journal={Journal of system simulation},
  volume={16},
  number={5},
  pages={893--897},
  year={2004}
}

@article{piao2025agentsociety,
  title={Agentsociety: Large-scale simulation of llm-driven generative agents advances understanding of human behaviors and society},
  author={Piao, Jinghua and Yan, Yuwei and Zhang, Jun and Li, Nian and Yan, Junbo and Lan, Xiaochong and Lu, Zhihong and Zheng, Zhiheng and Wang, Jing Yi and Zhou, Di and others},
  journal={arXiv preprint arXiv:2502.08691},
  year={2025}
}

@article{sivasankaran2014literature,
  title={Literature review of assembly line balancing problems},
  author={Sivasankaran, Panneerselvam and Shahabudeen, P},
  journal={The International Journal of Advanced Manufacturing Technology},
  volume={73},
  number={9},
  pages={1665--1694},
  year={2014},
  publisher={Springer}
}

@inproceedings{macal2009agent,
  title={Agent-based modeling and simulation},
  author={Macal, Charles M and North, Michael J},
  booktitle={Proceedings of the 2009 winter simulation conference (WSC)},
  pages={86--98},
  year={2009},
  organization={IEEE}
}

@inproceedings{macal2005tutorial,
  title={Tutorial on agent-based modeling and simulation},
  author={Macal, Charles M and North, Michael J},
  booktitle={Proceedings of the Winter Simulation Conference, 2005.},
  pages={14--pp},
  year={2005},
  organization={IEEE}
}

@article{gurcan2024llm,
  title={Llm-augmented agent-based modelling for social simulations: Challenges and opportunities},
  author={Gurcan, Onder},
  journal={arXiv preprint arXiv:2405.06700},
  year={2024}
}

@article{xue2024computational2,
  title={Computational experiments for complex social systems: Integrated design of experiment system},
  author={Xue, Xiao and Yu, Xiangning and Zhou, Deyu and Wang, Xiao and Bi, Chongke and Wang, Shufang and Wang, Fei-Yue},
  journal={IEEE/CAA Journal of Automatica Sinica},
  volume={11},
  number={5},
  pages={1175--1189},
  year={2024},
  publisher={IEEE}
}

@article{messeri2024artificial,
  title={Artificial intelligence and illusions of understanding in scientific research},
  author={Messeri, Lisa and Crockett, Molly J},
  journal={Nature},
  volume={627},
  number={8002},
  pages={49--58},
  year={2024},
  publisher={Nature Publishing Group UK London}
}

@misc{openai2024gpt4o,
  title        = {GPT-4o System Card},
  author       = {OpenAI},
  year         = {2024},
  url          = {https://openai.com/research/gpt-4o-system-card}
}

@misc{openai2025gpt5mini,
  title        = {GPT-5 Mini Model Card},
  author       = {OpenAI},
  year         = {2025},
  url          = {https://platform.openai.com/docs/models}
}

@inproceedings{ reimers-2019-sentence-bert,
Author = {Reimers, Nils and Gurevych, Iryna},
Book-Group-Author = {Assoc Computat Linguist},
Title = {Sentence-BERT: Sentence Embeddings using Siamese BERT-Networks},
Booktitle = {2019 CONFERENCE ON EMPIRICAL METHODS IN NATURAL LANGUAGE PROCESSING AND
   THE 9TH INTERNATIONAL JOINT CONFERENCE ON NATURAL LANGUAGE PROCESSING
   (EMNLP-IJCNLP 2019): PROCEEDINGS OF THE CONFERENCE},
Year = {2019},
Pages = {3982-3992},
Note = {Conference on Empirical Methods in Natural Language Processing / 9th
   International Joint Conference on Natural Language Processing
   (EMNLP-IJCNLP), Hong Kong, HONG KONG, NOV 03-07, 2019},
Organization = {Google; Facebook; Apple; ASAPP; Salesforce; Huawei; Baidu; Deepmind;
   Amazon; PolyAI; Naver; ByteDance; Megagon Labs; Zhuiyi; Verisk; MI},
ISBN = {978-1-950737-90-1},
Unique-ID = {WOS:000854193304015},
}

@article{zheng2023judging,
  title={Judging llm-as-a-judge with mt-bench and chatbot arena},
  author={Zheng, Lianmin and Chiang, Wei-Lin and Sheng, Ying and Zhuang, Siyuan and Wu, Zhanghao and Zhuang, Yonghao and Lin, Zi and Li, Zhuohan and Li, Dacheng and Xing, Eric and others},
  journal={Advances in neural information processing systems},
  volume={36},
  pages={46595--46623},
  year={2023}
}

@book{gilbert2019agent,
  title={Agent-based models},
  author={Gilbert, Nigel},
  year={2019},
  publisher={Sage Publications}
}

@article{jaeger2026decision,
  title={Decision Theaters and Democracy},
  author={Jaeger, Carlo and Laubichler, Manfred D},
  journal={Fairness and Competence in Citizen Participation: A Critical Review of Formats for Deliberative Policymaking},
  pages={97},
  year={2026},
  publisher={Springer Nature}
}

@article{huang2025survey,
  title={A survey on hallucination in large language models: Principles, taxonomy, challenges, and open questions},
  author={Huang, Lei and Yu, Weijiang and Ma, Weitao and Zhong, Weihong and Feng, Zhangyin and Wang, Haotian and Chen, Qianglong and Peng, Weihua and Feng, Xiaocheng and Qin, Bing and others},
  journal={ACM Transactions on Information Systems},
  volume={43},
  number={2},
  pages={1--55},
  year={2025},
  publisher={ACM New York, NY}
}

@article{chelli2024hallucination,
  title={Hallucination rates and reference accuracy of ChatGPT and bard for systematic reviews: comparative analysis},
  author={Chelli, Mika{\"e}l and Descamps, Jules and Lavou{\'e}, Vincent and Trojani, Christophe and Azar, Michel and Deckert, Marcel and Raynier, Jean-Luc and Clowez, Gilles and Boileau, Pascal and Ruetsch-Chelli, Caroline and others},
  journal={Journal of medical Internet research},
  volume={26},
  number={1},
  pages={e53164},
  year={2024},
  publisher={JMIR Publications Inc., Toronto, Canada}
}

@article{zhou2023sotopia,
  title={Sotopia: Interactive evaluation for social intelligence in language agents},
  author={Zhou, Xuhui and Zhu, Hao and Mathur, Leena and Zhang, Ruohong and Yu, Haofei and Qi, Zhengyang and Morency, Louis-Philippe and Bisk, Yonatan and Fried, Daniel and Neubig, Graham and others},
  journal={arXiv preprint arXiv:2310.11667},
  year={2023}
}

@inproceedings{zhou2025sotopia,
  title={SOTOPIA-S4: a user-friendly system for flexible, customizable, and large-scale social simulation},
  author={Zhou, Xuhui and Su, Zhe and Feng, Sophie and Zhou, Jiaxu and Huang, Jen-tse and Kao, Hsien-Te and Lynch, Spencer and Volkova, Svitlana and Wu, Tongshuang and Woolley, Anita and others},
  booktitle={Proceedings of the 2025 Conference of the Nations of the Americas Chapter of the Association for Computational Linguistics: Human Language Technologies (System Demonstrations)},
  pages={350--360},
  year={2025}
}

@article{gao2024large,
  title={Large language models empowered agent-based modeling and simulation: A survey and perspectives},
  author={Gao, Chen and Lan, Xiaochong and Li, Nian and Yuan, Yuan and Ding, Jingtao and Zhou, Zhilun and Xu, Fengli and Li, Yong},
  journal={Humanities and Social Sciences Communications},
  volume={11},
  number={1},
  pages={1--24},
  year={2024},
  publisher={Palgrave}
}

@misc{mao2025agent,
      title={Agent-Kernel: A MicroKernel Multi-Agent System Framework for Adaptive Social Simulation Powered by LLMs}, 
      author={Yuren Mao and Peigen Liu and Xinjian Wang and Rui Ding and Jing Miao and Hui Zou and Mingjie Qi and Wanxiang Luo and Longbin Lai and Kai Wang and Zhengping Qian and Peilun Yang and Yunjun Gao and Ying Zhang},
      year={2025},
      eprint={2512.01610},
      archivePrefix={arXiv},
      primaryClass={cs.MA},
      url={https://arxiv.org/abs/2512.01610}, 
}

@inproceedings{tang2025gensim,
  title={Gensim: A general social simulation platform with large language model based agents},
  author={Tang, Jiakai and Gao, Heyang and Pan, Xuchen and Wang, Lei and Tan, Haoran and Gao, Dawei and Chen, Yushuo and Chen, Xu and Lin, Yankai and Li, Yaliang and others},
  booktitle={Proceedings of the 2025 Conference of the Nations of the Americas Chapter of the Association for Computational Linguistics: Human Language Technologies (System Demonstrations)},
  pages={143--150},
  year={2025}
}

@misc{zhang2025socioverseworldmodelsocial,
      title={SocioVerse: A World Model for Social Simulation Powered by LLM Agents and A Pool of 10 Million Real-World Users}, 
      author={Xinnong Zhang and Jiayu Lin and Xinyi Mou and Shiyue Yang and Xiawei Liu and Libo Sun and Hanjia Lyu and Yihang Yang and Weihong Qi and Yue Chen and Guanying Li and Ling Yan and Yao Hu and Siming Chen and Yu Wang and Xuanjing Huang and Jiebo Luo and Shiping Tang and Libo Wu and Baohua Zhou and Zhongyu Wei},
      year={2025},
      eprint={2504.10157},
      archivePrefix={arXiv},
      primaryClass={cs.CL},
      url={https://arxiv.org/abs/2504.10157}, 
}

@article{xue2023chatgpt,
  title={ChatGPT chats on computational experiments: From interactive intelligence to imaginative intelligence for design of artificial societies and optimization of foundational models},
  author={Xue, Xiao and Yu, Xiangning and Wang, Fei-Yue},
  journal={IEEE/CAA Journal of Automatica Sinica},
  volume={10},
  number={6},
  pages={1357--1360},
  year={2023},
  publisher={IEEE}
}

@article{xue2023computational2,
  title={Computational experiments for complex social systems—Part III: the docking of domain models},
  author={Xue, Xiao and Yu, Xiangning and Zhou, Deyu and Peng, Chao and Wang, Xiao and Liu, Donghua and Wang, Fei-Yue},
  journal={IEEE Transactions on Computational Social Systems},
  volume={11},
  number={2},
  pages={1766--1780},
  year={2023},
  publisher={IEEE}
}

@article{xiao2023putational4,
  title={Com-putational experiments: Past, present and perspective},
  author={Xiao, Xue and Xiang-Ning, Yu and De-Yu, Zhou and Chao, Peng and Xiao, Wang and Zhang-Bing, Zhou and Fei-Yue, Wang},
  journal={Acta Automatica Sinica},
  volume={49},
  number={2},
  pages={246--271},
  year={2023}
}

@article{ma2024computational,
  title={Computational experiments meet large language model based agents: A survey and perspective},
  author={Ma, Qun and Xue, Xiao and Zhou, Deyu and Yu, Xiangning and Liu, Donghua and Zhang, Xuwen and Zhao, Zihan and Shen, Yifan and Ji, Peilin and Li, Juanjuan and others},
  journal={arXiv preprint arXiv:2402.00262},
  year={2024}
}

@article{xue2023computational5,
  title={Computational experiments: A new analysis method for cyber-physical-social systems},
  author={Xue, Xiao and Shen, Yifan and Yu, Xiangning and Zhou, De-Yu and Wang, Xiao and Wang, Gang and Wang, Fei-Yue},
  journal={IEEE Transactions on Systems, Man, and Cybernetics: Systems},
  volume={54},
  number={2},
  pages={813--826},
  year={2023},
  publisher={IEEE}
}

@article{xue2021soa,
  title={From soa to voa: a shift in understanding the operation and evolution of service ecosystem},
  author={Xue, Xiao and Zhou, Deyu and Chen, Fangyi and Yu, Xiangning and Feng, Zhiyong and Duan, Yucong and Meng, Lin and Zhang, Mu},
  journal={IEEE Transactions on Services Computing},
  volume={16},
  number={1},
  pages={315--329},
  year={2021},
  publisher={IEEE}
}

@article{peng2023computational,
  title={Computational experiments: Virtual production and governance tool for metaverse},
  author={Peng, Chao and Yu, Xiangning and Ma, Wanpeng and Kaneko, Hayata and Meng, Lin and Zhao, Yingyue and Xue, Xiao},
  journal={International Journal of Crowd Science},
  volume={7},
  number={4},
  pages={158--167},
  year={2023},
  publisher={TUP}
}

@article{yu2025causal,
  title={Causal sufficiency and necessity improves chain-of-thought reasoning},
  author={Yu, Xiangning and Wang, Zhuohan and Yang, Linyi and Li, Haoxuan and Liu, Anjie and Xue, Xiao and Wang, Jun and Yang, Mengyue},
  journal={arXiv preprint arXiv:2506.09853},
  year={2025}
}

@article{yu2025unlocking,
  title={Unlocking Complexity: Harnessing Value Entropy for Advanced Multidimensional Utility Evaluation in Service Ecosystems},
  author={Yu, Xiangning and Xue, Xiao and Zhou, Deyu and Wang, Gang and Feng, Zhiyong},
  journal={IEEE Transactions on Services Computing},
  year={2025},
  publisher={IEEE}
}

@inproceedings{yu2024beyond,
  title={Beyond traditional metrics: The power of value entropy in multidimensional evaluation of the service ecosystem},
  author={Yu, Xiangning and Xue, Xiao and Zhou, Deyu and Fang, Li and Feng, Zhiyong},
  booktitle={2024 IEEE International Conference on Web Services (ICWS)},
  pages={611--621},
  year={2024},
  organization={IEEE}
}
